\documentstyle[preprint,aps]{revtex}

\begin{document}

\draft

\title{Interactions of Ar$^{9+}$ and metastable Ar$^{8+}$ with a
  Si(100) surface at velocities near the image acceleration limit}

\author{J.~Ducr\'ee\thanks{author to whom correspondence should be
    addressed.Electronic address: ducree@uni-muenster.de},
  J.~Mrogenda, E.~Reckels, M.~R\"uther, A.~Heinen, Ch.~Vitt,
  M.~Venier, J.~Leuker, and H.J.~Andr\"a}

\address{Institut f\"ur Kernphysik, Westf\"alische
  Wilhelms-Universit\"at M\"unster, Wilhelm-Klemm-Str.~9, D-48149
  M\"unster, Germany}

\author{R.~D\'{\i}ez Mui\~{n}o} 
\address{Laboratoire de Physico-Chimie Th\'eorique, C.N.R.S. et
  Universit\'e de Bordeaux I, 351 Cours de Lib\'eration, 33405 Talence
  Cedex, France}

\date{\today}

\maketitle

\begin{abstract}
  Auger $LMM$ spectra and preliminary model simulations of Ar$^{9+}$ and
  metastable Ar$^{8+}$ ions interacting with a clean monocrystalline
  $n$-doped Si(100) surface are presented.  By varying the
  experimental parameters, several spectroscopic features have been
  observed providing valuable information for the development of an
  adequate interaction model.  On our apparatus the ion beam energy
  can be lowered to almost mere image charge attraction. High data
  acquisition rates could still be maintained yielding an
  unprecedented statistical quality of the Auger spectra.
\end{abstract}


\section{Introduction}
\label{sec:intro}

The interactions of highly charged ions (HCI) with surfaces have
attracted the strong interest of several research groups in the past,
getting a strong boost in the last decade due to the increasing
availability of high performance HCI ion sources and improvement of
other experimental equipment.

In recent years, various technological applications of HCI surface
collisions have been conceived, in particular for the wide field of
microscopic and nanoscopic surface modification. In order to foster
these efforts, a better understanding of the different stages of the
scattering process has to be attained. Experimentalists hope to take
advantage of the quick charge exchange processes and the release of
the large amount of potential energy stored in the HCI.  Unfortunately
little consent has been accomplished among researchers on the time
scales and the location of these processes although a comprehensive
series of spectra and interpretations has been published on this
crucial issue so far.

According to the classical overbarrier model \cite{Bur91,Duc97}, the
neutralization of the HCI sets in at a critical distance of typically
$R_c \simeq 15$ {\AA} in front of the first bulk layer. $R_c$ depends
on the target work function $W$ and the initial charge $q$ of the HCI.
In the region below $R_c$, target band electrons are successively
captured into resonant ionic Rydberg states with $n \simeq q
\sqrt{R_\infty/W}$.  As soon as more than two electrons have been
transferred, the highly excited hollow atom starts to relax via
autoionization processes yielding low-energy electrons.  X-ray
emission is strongly suppressed for light nuclei. Several studies
\cite{Mey95,Hat96} have been carried out showing that the overwhelming
fraction of the reflected particles is neutral and suggesting that the
projectile charge $q$ is already compensated on the incoming path.
Nevertheless, it is commonly accepted by now \cite{Sch94} that the
intra-atomic transition rates involved in the cascade are by far too
slow to perform a complete relaxation of the neutralized HCI in front
of the surface.

Autoionization spectra originating from highly charged ions containing
initial inner-shell vacancies are characterized by a strong and
intense low-energy region and a uniquely shaped high-energy branch
which can unambiguously be ascribed to intra-atomic transitions
involving the inner-shell vacancy.  Despite of the low transition
rates, certain peak structures can even be associated with Auger
emission from fully populated shells neighboring the initial core
configuration.

In order to clarify the evolution from Rydberg populations to fully
occupied lower shells and motivated by new experimental findings
\cite{Mey91,Koe94,Hst94} about large fractions of subsurface emission
within the autoionization spectra, additional interaction mechanisms
have been postulated and worked into simulations
\cite{Lim95a,Pag95,Sto95}. Also a comparison between Auger spectra for
the same HCI projectile impinging on different target species
\cite{Lim96} and a new theoretical approach \cite{Arn95a} shed new
light on the interaction scenario. It seems that the energetic
positions of target and projectile electronic states play an important
role in all direct inner-shell filling mechanisms below the surface.
After the HCI has penetrated into the bulk, band electrons can shield
the HCI core charge and directly feed the lower lying hollow atom
states while generating a plasmon or an electron hole pair
\cite{Die95,Die96} (so called $MCV$ processes).  For projectiles
with high kinetic energies, electrons can be directly transferred from
bulk atom levels into inner projectile levels yielding a velocity
dependent filling rate \cite{Lim95a}.

In an attempt to extract information on particular transition types
from the spectra, experimentalists have analyzed $L$-Auger spectra of
Ar$^{9+}$ ions impinging on tungsten \cite{Zwa87,Fol89,Zwa89,And91A},
copper \cite{Koe92} and gold \cite{Mey91}. These early efforts have
been obstructed by the large number of initial $M$-shell configurations
that had to be considered in the interpretation of the $LMM$ spectra
with a few distinctive structures only.  In recent years, research
activities focused on $K$-Auger spectra of hydrogenlike second row ions
C$^{5+}$, N$^{6+}$, O$^{7+}$, F$^{8+}$, Ne$^{9+}$
\cite{Hst94,And93,Fol90,Lim94a} and Ar$^{17+}$ \cite{Mey95} instead.
Some clearly pronounced peak regions can be identified in most of
these spectra and assigned to a comparatively small set of initial
$L$-shell configurations. A strong systematic dependence of the
relative peak intensities of these $KLL$ spectra on the experimental
conditions has provided valuable information about the contributing
ionic shell configurations.

In this paper, we present several series of $L$-Auger spectra emitted
during the interaction of Ar$^{9+}$ and metastable ``heliumlike''
Ar$^{8+}$ ions impinging with beam energies between 8eV and 4.6~keV and
different experimental geometries on an n-doped Si(100) crystal.  For
the first time, we discovered significant modifications of the shape
of the autoionization spectra for different projectile energies well
below 1~keV and for different observation and interaction geometries.
They include two geometries that largely suppress the detection of
all subsurface electron emission.  The so obtained spectra exhibit a
unique peak profile that largely deviates from spectra taken under
all other experimental geometries.

These effects are very surprising because in the energy regime below
1~keV all collisional $M$-shell sidefeeding can generally be ruled out
and $MCV$ rates can be treated in a static approximation.  In order to
understand the behavior of the spectra at different incident energies
we developed an interaction model taking into account the special role
of the $3d$ subshell, which mediates an efficient $M$-shell filling via
valence band electrons within the bulk.  Incorporating this model into
a Monte Carlo simulation, the observed alterations in the subpeak
intensities and positions can qualitatively be reproduced.  This model
is experimentally supported by a series of $L$-Auger spectra emitted by
metastable Ar$^{8+}$ projectiles. Under the same experimental
conditions, the Ar$^{8+}$ $LMM$ Auger peak structures turn out to be
amazingly similar to the Ar$^{9+}$ $LMM$ structures.

Section~\ref{sec:setup} will introduce to the experimental setup
implemented for our measurements.  In Section~\ref{sec:observations},
we display several sets of autoionization spectra as obtained under
specified experimental conditions.  Section~\ref{sec:grouping} will
describe how the $LMM$ subpeaks in our Ar$^{9+}$ spectra can be
assigned to particular groups of intra-atomic transitions.  The next
Section~\ref{sec:subsim} outlines the basic ingredients of the
subsurface interaction model which we employ for the simulation of the
Auger spectra.  In Section~\ref{sec:evolution}, we extract information
about the evolution of the projectile neutralization and Auger
emission from the combined analysis of experimental observations and
the simulation results. Further experimental proof for the portrayed
interaction mechanism will be given with the discussion of the {\it
  L\/}-Auger spectra of metastable Ar$^{8+}$ projectiles
in~\ref{sec:Ar8spectra}.  Finally, in~\ref{sec:discussion}, we
summarize the basic findings of this paper and give a short outlook on
future research.

\section{Experimental setup}
\label{sec:setup}

Highly charged ions are extracted by a fixed voltage of --20kV from an
ECR ion source, developed in our laboratory. The metallic vacuum
chamber of the source can be floated on selectable potentials $U_Q$
with respect to earth potential. These ions are $q/m$ separated by a
double focusing sector magnet system including an aberration
correction lens.  Two electrostatic Einzel lenses convey the beam
through the intermediate stages of a differentially pumped vacuum
system which is needed to maintain the pressure gradient between the
ECR source ($p \simeq 1 \times 10^{-6}$mbar) and the UHV target
chamber ($p \simeq 5 \times 10^{-12}$mbar). Before hitting the grounded
Si wafer, the ions pass through two deceleration lenses which are
optimized for a maximum of ions deposited on the target surface of
approximately 1cm$^2$.

The kinetic ion energy distribution is recorded by an ion spectrometer
which is mounted on the beam axis close behind the movable target.
For Ar$^{9+}$ and Ar$^{8+}$ beams, the full width at half maximum
never exceeded 2eV per charge. The center of the peak is a measure for
the kinetic projectile energy after deceleration $E_{\mbox{kin}} = q
(U_Q + U_P)$ where $U_P$ is the plasma potential which builds up
between the plasma and the walls of the ECR source.  An averaged value
of $U_P = 12$V has been observed with variations over months of less
than $\pm$2V.  The Si(100) surface has been prepared by successive
cycles of Ar$^{1+}$ sputtering at grazing incidence and annealing
until all impurities have disappeared from AES spectra and good LEED
patterns have shown up.

The geometry within the target chamber is displayed in
Fig.~\ref{fig:geometry}(a). The beam axis intersects the target
surface at an angle $\Theta$. Electrons are detected by an
electrostatic entrance lens followed by a $150^\circ$ spherical sector
analyzer at an angle $\Psi$ with respect to the surface. In most
measurements we chose $\Theta+\Psi=90^\circ$. As $\Psi$ approaches
$0^\circ$ in Fig.~\ref{fig:geometry}(b), the path length inside the
solid for electrons which are emitted below the surface drastically
increases such that the detection of above or near surface emission is
clearly favored. Due to the chamber alignment and the large acceptance
angle of $\eta = 16\pm6^{\circ}$ of our electron spectrometer entrance
lens, below-surface emission is always observed, but to a much
smaller extent than above or near surface emission.  The absolute
spectral intensity in the ($\Psi \simeq 0^\circ$)-geometries greatly
diminishes, though. By rotating the target of
Fig.~\ref{fig:geometry}(a) around the ion beam axis with the surface
normal pointing out of the image plane, the condition of
$\Theta+\Psi=90^\circ$ could be relaxed, and geometries with
$\Theta=5^\circ$ and $\Psi=0^\circ$ have been achieved.

The effective incident energy of the ions on the surface is given by
$E_{\mbox{kin}}$ plus the energy gain resulting from the image charge
acceleration \cite{Win93}
\begin{equation}
  E_{\mbox{im}} \simeq \frac{W}{3\sqrt{2}} q^{3/2} 
\end{equation}
where the work function $W$ equals 4.6eV for our Si target and $q =
9$. Accordingly, there will always remain a minimum incident energy of
approx.\ 29eV leading to an additional perpendicular projectile
velocity component $\Delta v_{\perp} = \sqrt{2 \cdot
  E_{\mbox{im}}/m}$. Thus the interaction period of the ion in front
of the surface can principally not be stretched above an upper limit
depending on $q$ and $W$ even though the original perpendicular
velocity component $v_{\perp} = \sqrt{2 E_{\mbox{kin}}/m} \cdot
\cos(\Theta)$ of the projectile may vanish by selecting $U_Q = -U_P$
or $\Theta \mapsto 0^\circ$.

When the incident energy $E_{\mbox{kin}}$ is lowered the beam spreads
up at the target (Liouville's theorem) and incident angles may deviate
from their nominal values $\Theta$. In the energy domain
$E_{\mbox{kin}} < E_{\mbox{im}}$, the projectile path is strongly bent
by the attractive image acceleration causing increased effective
incident angles $\Theta_{\mbox{eff}}$, especially for small $\Theta$.
Hence the values given for $\Theta$ in this paper are intended to
delineate the chamber geometry rather than the effective scattering
geometry of an individual projectile.

Projectile penetration depths at the stage of complete neutralization
and deexcitation can be estimated by multiplying $v_\perp$ by a
typical overall interaction time of $10^{-14}$s. With $E_\perp =
\frac{1}{2} m v_\perp^2$ expressed in eV, the perpendicular path
length $z_{pen}$ of the Ar projectile within the bulk can be attained
from $z_{pen} = 0.22$\AA$\times \sqrt{E_\perp [\mbox{eV}]}$. This
implies that at energies $E_\perp$ in the range of 100eV, $z_{pen}$
stays below one lattice constant amounting to 5.43\AA\ for Si.
{\sc TRIM} simulations \cite{Zie85} performed for a 10eV and a 100eV
Ar$^{1+}$ beam impinging on a Si crystal at perpendicular incidence
yield average lateral ranges of 3$\pm$1\AA and 10$\pm$4\AA,
respectively. These distances refer to the total penetration depth
until the ion is stopped within the bulk.

\section{Experimental observations}
\label{sec:observations}

Using the apparatus described in the preceeding section, we have
measured secondary electron spectra emitted by Ar$^{9+}$ and
metastable Ar$^{8+}$ ions during their interaction with the Si wafer.
In this work we will focus on examining the well defined high-energy
$L$-Auger peaks covering the interval between 120eV and 300eV.  The
spectra also feature a low-energy part which extends up to more than
100eV.  The analysis of electron spectra in this energy domain is
aggravated by a lack of substructures, the superposition of kinetic
and intra-atomic emission and their sensitivity to stray
electromagnetic fields.  Regarding the high-energy branch, we point
out that no background due to kinetic electron emission has to be
considered for $E_{\mbox{kin}} \leq 121$~eV since the collision energies
$E_{coll}=E_{\mbox{kin}}+E_{\mbox{im}}$ are smaller than the lower bound of the
spectral region to be examined. By selecting $U_Q=-20$V$<U_P= 12 \pm
2$V, we can prevent HCIs from reaching the grounded target. Only
projectiles that are partially neutralized before the deceleration
stages and secondary electrons which are generated by collisions of
the HCIs with beam transport lens elements (these are on negative
potentials) can hit the target where they may set free secondary
electrons. We discovered that both contributions are negligible.

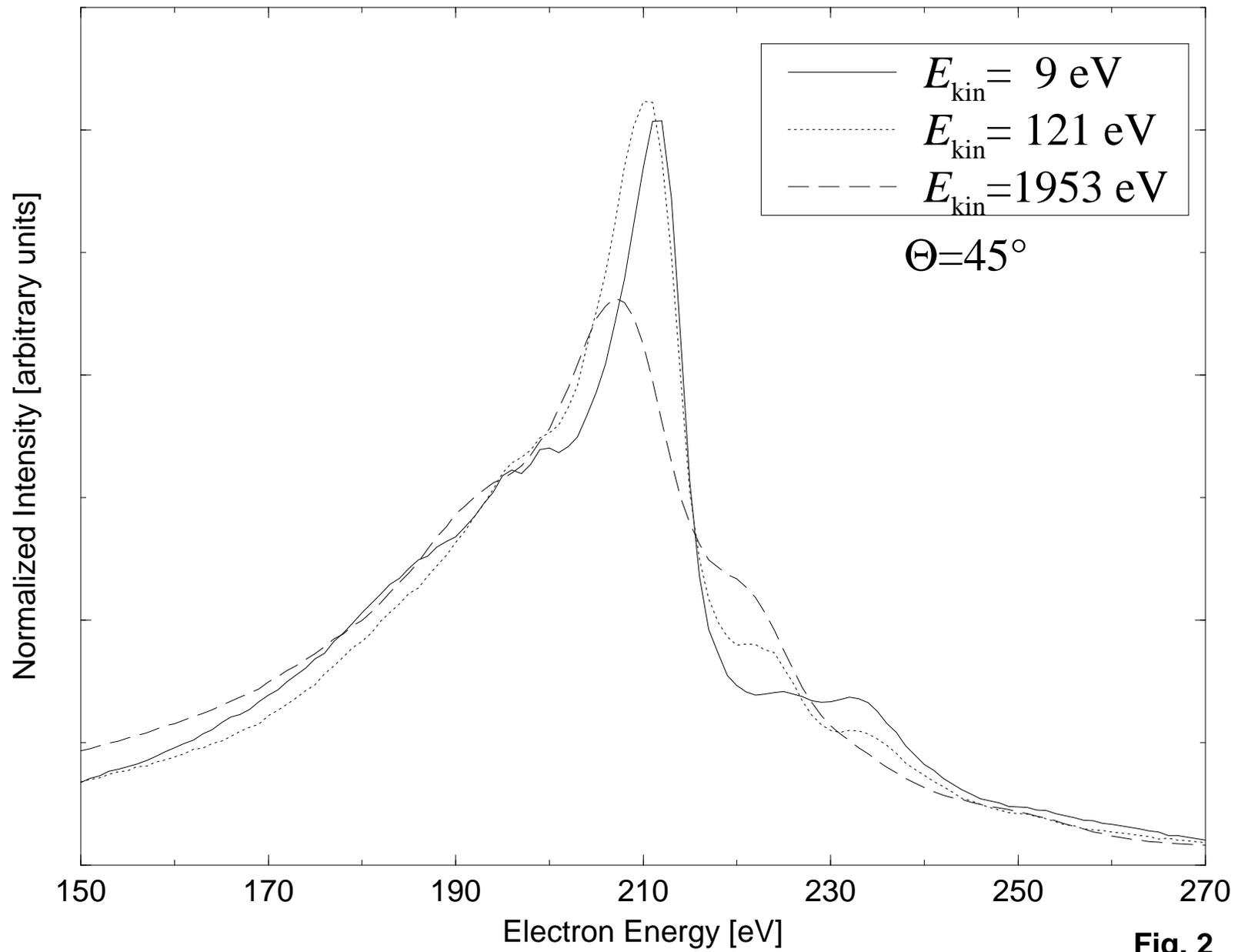
\begin{figure}[p]  
  \caption{Experimental spectra of Ar$^{9+}$ projectiles impinging
    at a fixed incident angle of $\Theta=45^\circ$ and various
    energies on a $n$-Si(100) surface. All spectra in this paper are
    normalized on the portrayed peak region between 160eV and 240eV.
    The error at the peak maximum of the
    ($E_{\mbox{kin}}=121$~eV)-spectrum is smaller than 0.3\%. With
    increasing incident energy $E_{\mbox{kin}}$, the 224eV-peak begins
    to outgrow the 232eV-peak and the main peak moves from 211eV to
    208eV.}
  \label{fig:Ar9Si:45deg:energy:normreg}
\end{figure}

In Fig.~\ref{fig:Ar9Si:45deg:energy:normreg} we present three
Ar$^{9+}$ spectra measured under $\Theta=45^\circ$ and with
$E_{\mbox{kin}}=9$~eV, 121eV and 1953eV. This and all following spectra are
normalized to the total intensity in the $L$-Auger region between 160eV
and 240eV. 
Considering that at maximum one $L$-Auger process per ion takes place,
this type of normalization method is suitable to display the intensity
shifts between $L$-transition subgroups as discussed in this paper. We
note that the calibration of the spectra to the absolute beam
intensity is prone to errors which emerge from the uncertainty in the
correction factors compensating geometrical and kinetic effects.

At first we recognize the general shape of an Ar$^{9+}$ $LMM$ spectrum
featuring a dominant peak at 211eV, a broad structure reaching down to
about 120eV on the low-energy side and a shoulder sitting on the high
energy tail of the spectrum. At $E_{\mbox{kin}}=9$~eV, this shoulder can be
resolved into two subpeaks of almost equal height at 224eV and 232eV.
Proceeding to higher $E_{\mbox{kin}}$, the 232eV-peak disappears and the
224eV-peak gains intensity. Presumably due to the poor statistical
quality, the latter 232eV-substructure cannot unambiguously be
identified in de Zwart´s \cite{Zwa89} measurement\footnote{There has
  obviously been a mistake in the calibration of the plot on the
  energy axis that has been corrected in \cite{Fol89}.} which was
taken under the same experimental geometry and roughly the same
incident energy on a tungsten target.

The data acquisition statistics of our spectra exhibits a remarkably
high quality. Beam current shifts during measurements are compensated
by an online normalization of the spectra to the overall charge
current $I_q$ hitting the target.  The accumulated counts per 1eV
energy channel in the ($\Theta=45^\circ$,$E_{\mbox{kin}}=121$~eV)-spectrum
amount to more than 200,000 at the 211eV-maximum letting the relative
error drop below 0.3\%. We note that each spectrum in
Fig.~\ref{fig:Ar9Si:45deg:energy:normreg} has been recorded in a
single five minute run. This is possible due to the high current
$I_q=125$nA on the target which can be converted into a particle
current $I_p$ by dividing $I_q$ over the projectile charge $q$ and
applying a correction factor compensating secondary electron emission.
Multiplying $I_p$ by an appropriate geometrical factor, it can be
shown that the overall experimental count rate in the high-energy
branch roughly correlates to the emission of one high-energy electron
per incoming HCI.

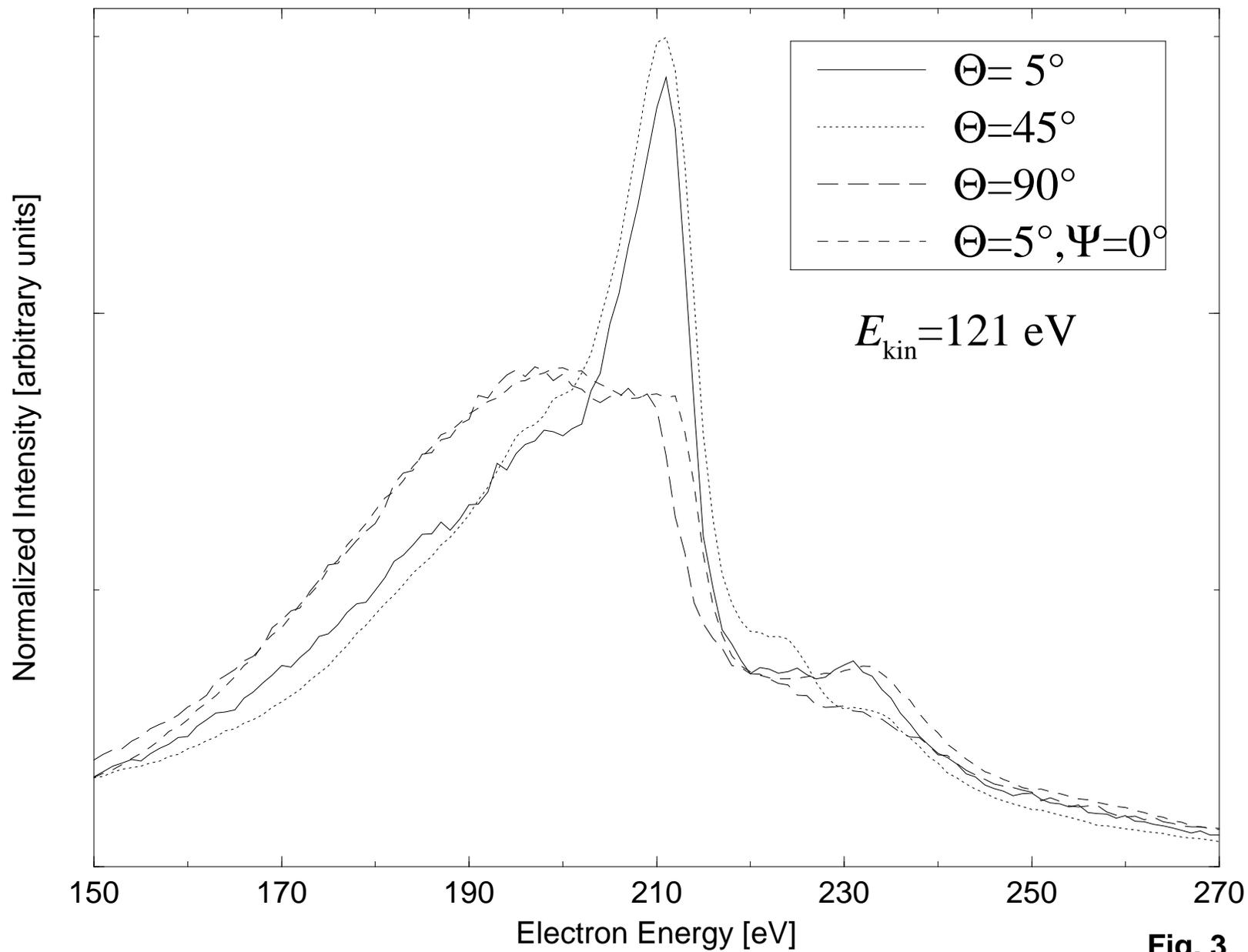
\begin{figure}[p]  
  \caption{Ar$^{9+}$ spectra at $E_{\mbox{kin}}=121$~eV and
    varying incident angles $\Theta$. The intensities of the two
    subpeaks on the shoulder above 211eV alter with $v_\perp$ in the
    same way as in Fig.~\ref{fig:Ar9Si:45deg:energy:normreg}.  For the
    two geometries at $\Psi=0^\circ$, a second broad peak region
    evolves around 198eV which we associate with emission from
    partially ionized incoming or reflected particles.}
  \label{fig:Ar9Si:121eV:angle:normreg}
\end{figure}

The spectral series of Ar$^{9+}$ ions impinging on Si(100) with
constant $E_{\mbox{kin}}=121$~eV in Fig.~\ref{fig:Ar9Si:121eV:angle:normreg}
displays the variation of the relative peak intensities with the
experimental geometry. Recalling
Fig.~\ref{fig:Ar9Si:45deg:energy:normreg}, we discover that the
presence of a strong 232eV-subpeak is connected to minimum
perpendicular velocities $v_\perp$.  In the measurement under
$\Theta=90^\circ$, the observation angle $\Psi$ is very flat and a
second broad peak region evolves around 198eV.  Switching to the other
``grazing observation'' alignment at $\Theta=5^\circ,\Psi=0^\circ$,
this structure is preserved proving that its presence is related to a
small observation angle $\Psi$ rather than the direction of incidence
$\Theta$ or $\Theta_{\mbox{eff}}$.

Under $\Theta=5^\circ$, the perpendicular projectile penetration into
the bulk is principally limited to less than one lattice constant. The
severe discrepancy between the two spectra under $\Theta=5^\circ$ and
$\Theta=5^\circ,\Psi=0^\circ$ in
Fig.~\ref{fig:Ar9Si:121eV:angle:normreg} illustrates the extreme
above-surface sensitivity of the ($\Psi=0^\circ$)-measurements since
the physics of interaction is only determined by $\Theta_{\mbox{eff}}$ and
$E_{\mbox{kin}}$ which remain constant.  We deduce that the broad peak region
is generated above or at least near the first bulk layer.  Because
this region looses its weight under $\Psi=45^\circ$ when electrons
originating from all interaction phases are detected, above-surface
processes only supply a minor fraction of the total high-energy
emission. Nevertheless, the ratio between the $detected$ above-
and below-surface emission is strongly enhanced at grazing observation
$\Psi = 0^\circ$ and small projectile penetration depths.

To obtain a quantitative estimate, we ran \textsc{TRIM} calculations
\cite{Zie85} for Ar$^{1+}$ ions colliding with a Si target. The
results show that a few percent of the incoming particles are
reflected for $E_{\mbox{kin}}=121$~eV and 2~keV complying with the preceeding
interpretation of the ($\Psi = 0^\circ$)-spectra. We point out that
one has to be careful about adopting these findings for HCI beams
because the \textsc{TRIM} code solely employs potentials which are
strictly speaking only valid for onefold ionized ground state
projectiles. For incident energies of less than 10eV when $E_{\mbox{im}} >
E_{\mbox{kin}}$, the code fails to produce physically meaningful output since
it obviously misrepresents the potentials evolving from the complex
coupling of the HCI-surface system. These potentials are decisive for
the calculation of the HCI trajectory along the prolonged interaction
period in front of the surface and the reflection probability. At such
low incident energies, no experimental data on reflection coefficients
of Ar$^{q+}$ impinging on Si(100) are available in the literature or
refers to grazing incidence conditions where the physics of
interaction is different despite the similar vertical velocity
components. The detection of the unique peak profile at grazing
observation combined with the oncoming discussion may be regarded as
indirect experimental evidence for the existence of reflected
projectiles.

The shifts of the upper edge of the 211eV-peak in
Fig.~\ref{fig:Ar9Si:121eV:angle:normreg} can consistently be explained
by an enhanced below-surface damping of the emitted electrons at
$\Theta=90^\circ$ which is more effective than at
$\Theta=5^\circ,\Psi=0^\circ$ due to the higher perpendicular velocity
component $v_\perp$.

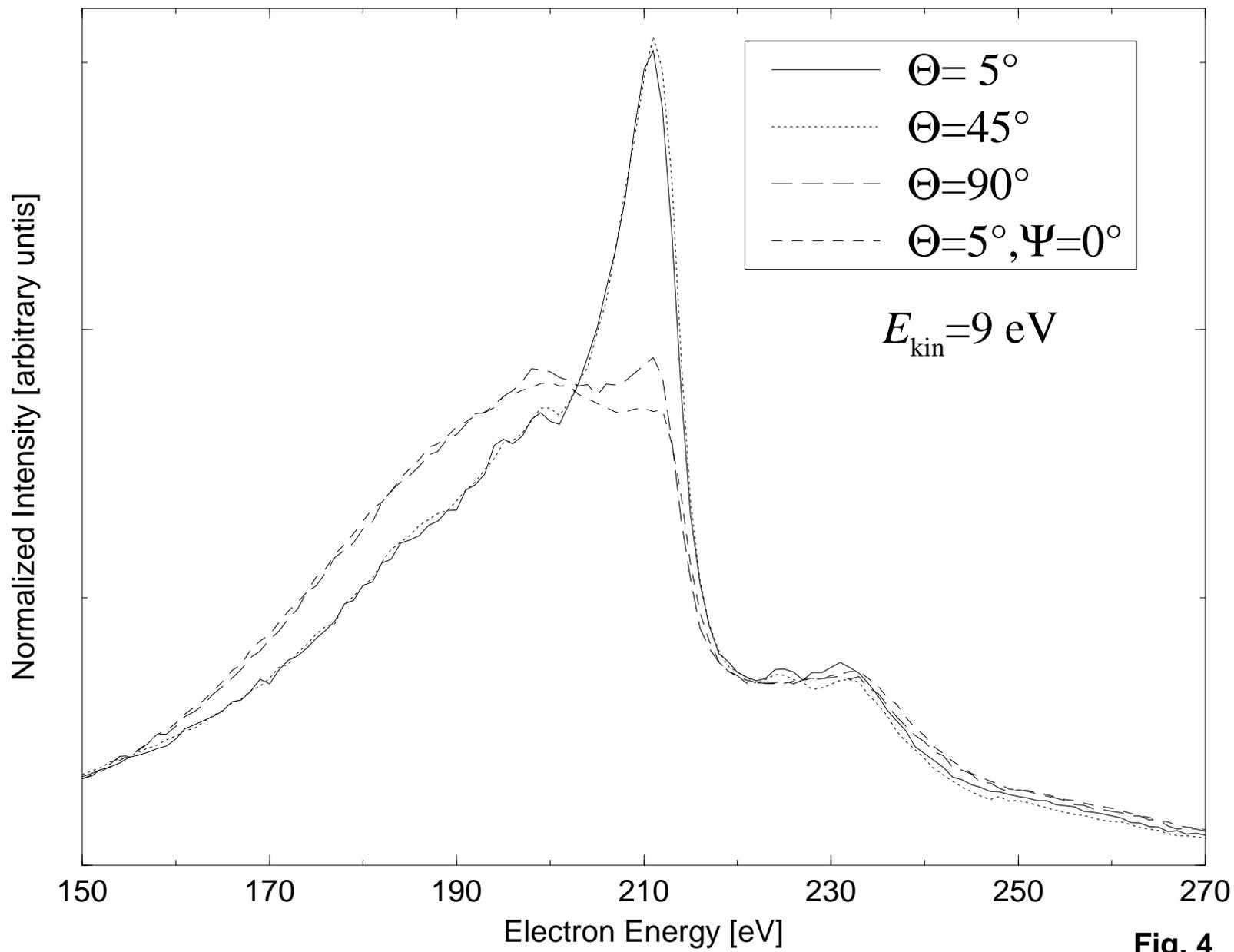
\begin{figure}[p]  
  \caption{Ar$^{9+}$ spectra at  $E_{\mbox{kin}}=9$~eV and
    $\Theta$ as parameter. The perpendicular velocity component
    $v_\perp$ is minimal in all four measurements and
    $\Theta_{\mbox{eff}}$ thus stays roughly constant due to the image
    potential. Due to our normalization method, the high-energy
    shoulder above 211eV is untouched by changing the nominal incident
    angle $\Theta$. At $\Psi = 0^\circ$, the broad region around 198eV
    pops up again.}
  \label{fig:Ar9Si:9eV:angle:normreg}
\end{figure}

In Fig.~\ref{fig:Ar9Si:9eV:angle:normreg} we show spectra of Ar$^{9+}$
ions impinging on a n-Si(100) surface under different incident angles
with minimal kinetic energies, i.e., $E_{\mbox{kin}} = 9$~eV. For
$\Theta=5^\circ$ and $\Theta=45^\circ$, the spectra are nearly
identical reflecting the fact that the self-image attraction is
greater than the kinetic projectile energy so that the effective angle
of incidence $\Theta_{\mbox{eff}}$ becomes almost independent of its original
value $\Theta$. While approaching perpendicular incidence, the same
broad region between 160eV and 205eV as in
Fig.~\ref{fig:Ar9Si:121eV:angle:normreg} pops up again. For the two
different ($\Psi=0^\circ$)-geometries, the main peaks exhibit about
the same height. Since $v_\perp$ is minimal in all four spectra, the
upper edge of the 211eV-peak remains sharp and does not shift to
lower energies due to bulk damping as in
Fig.~\ref{fig:Ar9Si:45deg:energy:normreg}. 
Even more, the high-energy branches above 211eV coincide almost
perfectly. Keeping in mind our particular choice of normalization
method and the minimum incident energy $E_{\mbox{kin}} = 9$~eV, the latter
feature suggests that the peak intensity within the high-energy tail
region results from above-surface emission which is insensitive to
bulk damping of the outgoing electrons.

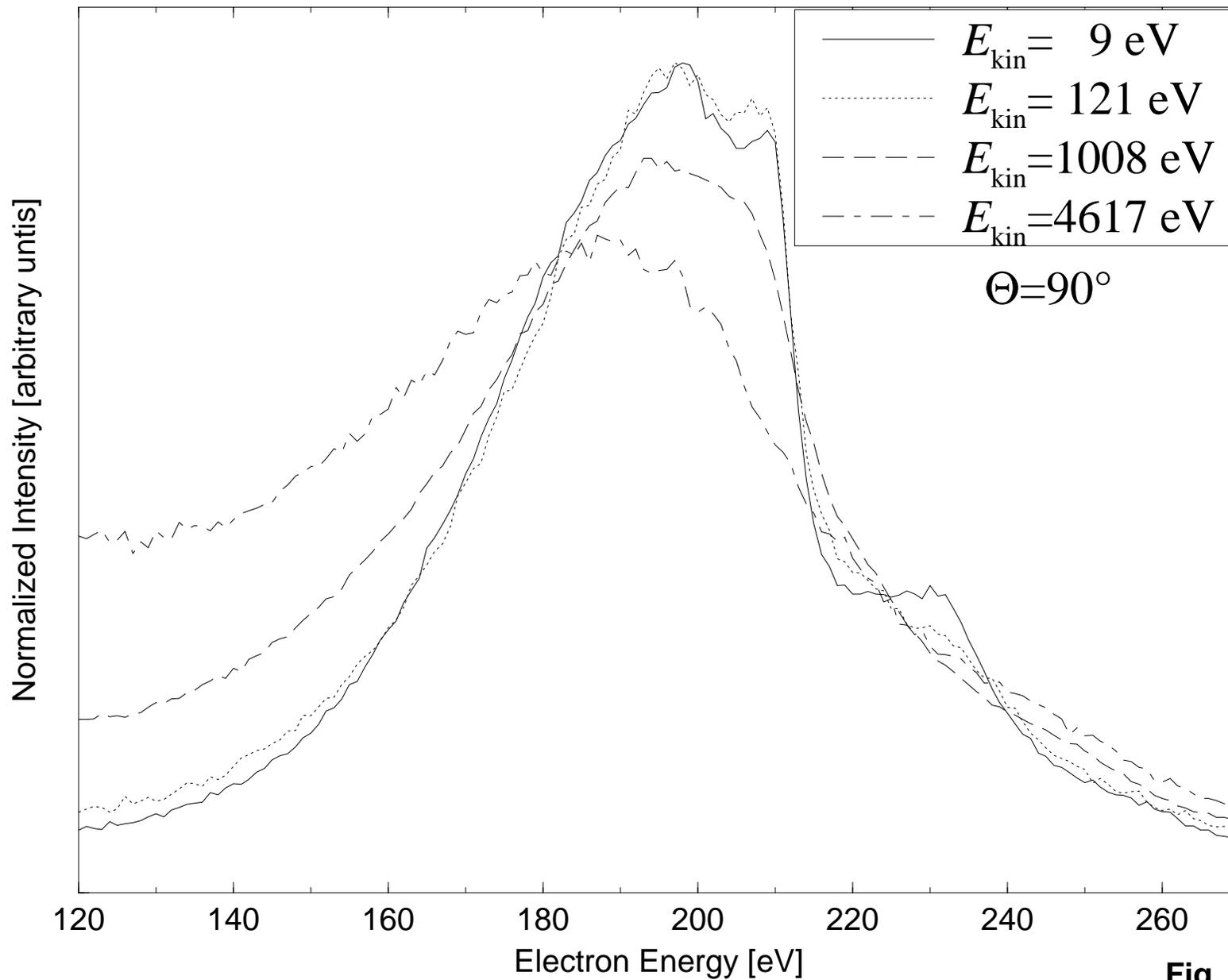
\begin{figure}[p]  
  \caption{Series of Ar$^{9+}$ spectra taken at fixed angles
    $\Theta=90^\circ$ and $\Psi=0^\circ$ with varying
    $E_{\mbox{kin}}$. For $E_{\mbox{kin}}=9$~eV and 121eV, we can
    still distinguish two separate peaks around 198eV and 211eV.  For
    higher $E_{\mbox{kin}}$, the 211eV-peak disappears and the broad
    structure washes out while extending to lower energies. The
    232eV-peak can only be found for minimum $v_\perp$.}
  \label{fig:Ar9Si:92deg:energy:normreg}
\end{figure}

In Fig.~\ref{fig:Ar9Si:92deg:energy:normreg} we present another series
of Ar$^{9+}$ spectra taken at a fixed angle $\Theta=90^\circ$ (i.e., 
$\Psi=0^\circ$) for different incident energies $E_{\mbox{kin}}$. As the
point of emissions moves deeper into the solid, below-surface
contributions are successively filtered out by bulk damping. The
double-peak profile transforms into a single unstructured maximum
widening to the low-energy side as $E_{\mbox{kin}}$ increases. The low-energy
bounds of the 198eV maximum coincide at $E_{\mbox{kin}}=9$~eV and 121eV.  The
spectrum measured at $E_{\mbox{kin}}=121$~eV demonstrates that the appearance
of the broad peak structure under $\Psi = 0^\circ$ and the 232eV-peak
occurring solely at minimal $v_\perp$ are obviously not immediately
linked to each other.

The combined analysis of the spectra in
Figs.~\ref{fig:Ar9Si:45deg:energy:normreg}-\ref{fig:Ar9Si:92deg:energy:normreg}
renders the following preliminary picture which will be supported by
further evidence and simulations in the next sections.  The dominant
211eV-peak originates from below-surface emission since its center
moves downward, it broadens and its intensity decreases when long path
length of the emitted electrons through the bulk to the spectrometer
entrance can be assumed.  Furthermore, it does not disappear with
growing $v_\perp$.  This also holds for the lower lying part of the
spectrum.  Two equally intense subpeaks on the high-energy shoulder
exclusively appear when $v_\perp$ is minimized.  As $v_\perp$
increases, the 224eV-peak gains intensity while the 232eV-peak
quickly vanishes.  This behavior suggests a dependence of the
232eV-intensity on the above-surface interaction time even though
the resulting emission process may occur after surface penetration.

The broad peak region between 160eV and 205eV under $\Psi=0^\circ$ and
$E_{\mbox{kin}} \leq 121$~eV represents near- or above-surface emission
since the ``detection window'' is shallow and the chamber geometry
favors detection of above-surface transitions at the same time.
Subsurface contributions are shielded by bulk damping.  For reasons
that will be given in Section~\ref{sec:Ar8spectra}, it is likely that
it is made up of a small fraction of above-surface emission from
partially screened incoming or ionized reflected particles. The
preceeding experimental findings will play a crucial role in the
conception of an interaction model in Section~\ref{sec:evolution}.

\section{Energetic grouping of atomic $LMM$ transitions}
\label{sec:grouping}

In this section we will attribute some spectral features occurring in
the energy range between 150eV and 300eV to distinct groups of $LMM$
Auger transitions. The energetic overlap between neighboring groups
will ``fortunately'' turn out to be sufficiently small such that
relative peak intensities can be related to the participation of
distinguished Auger processes.  Furthermore, certain projectile
deexcitation mechanisms can definitively be ruled out if no intensity
is measured in their proper energy range. By merely comparing peak
energies, we obtain valuable information concerning the HCI-solid
interaction which supplement the experimental observations of
Section~\ref{sec:observations} \textit{before} launching any
simulation. At the present state of research, peak energies can be
evaluated more accurately than transition rates for the HCI solid
system.

We employ the well known Cowan code \cite{Cow81} in order to simulate
configuration energies based on spherically symmetrized wave functions
for \textit{free} atoms and ions. In order to calculate Auger
transition energies \textit{within the bulk}, we have to take into
account the effect of the self induced charge cloud consisting of
valence band (VB) electrons which surrounds the HCI. First approaches
have been made on this behalf \cite{Arn95a,Arn95} using the density
functional theory (DFT).  Results show that the nonlinear screening
effects due to the electron gas are to a good approximation equivalent
to the screening by outer shell ``spectator'' electrons in a free
atom.

The hollow atom entering the bulk loses all Rydberg shell electrons
due to the screening by the target electron gas.  The radii of the
resonantly populated orbitals are of the order of the capture
distances, i.e., about 10{\AA} and therefore much larger than the
Thomas-Fermi screening length of less than one {\AA}ngstr{\o}m as
derived in a free electron gas model. Therefore all Rydberg levels
will be depleted leaving behind the original $1s^22s^2p^5$ core
configuration and possibly some $M$-, $N$- and $O$-shell electrons. The
target electron gas will swiftly take over the role of the outer
electrons to screen and so neutralize the HCI charge. A good estimate
for the reaction rate of the electron gas to the HCI ``point charge''
perturbation is provided by the plasmon frequency which lies in the
vicinity of $10^{16}$s$^{-1}$ for metals.  This is way above typical
rates of the other HCI bulk interaction processes and we can thus
assume that the HCI core screening by VB electrons is instantaneous.
Except for the special handling of the transitions with
$3d$ participation, which will be outlined below, all subsurface
Auger transition energies given in this paper will hence be derived
for neutral initial states possessing a total amount of $q$ $M$- and
$N$-shell electrons and singly ionized final states.

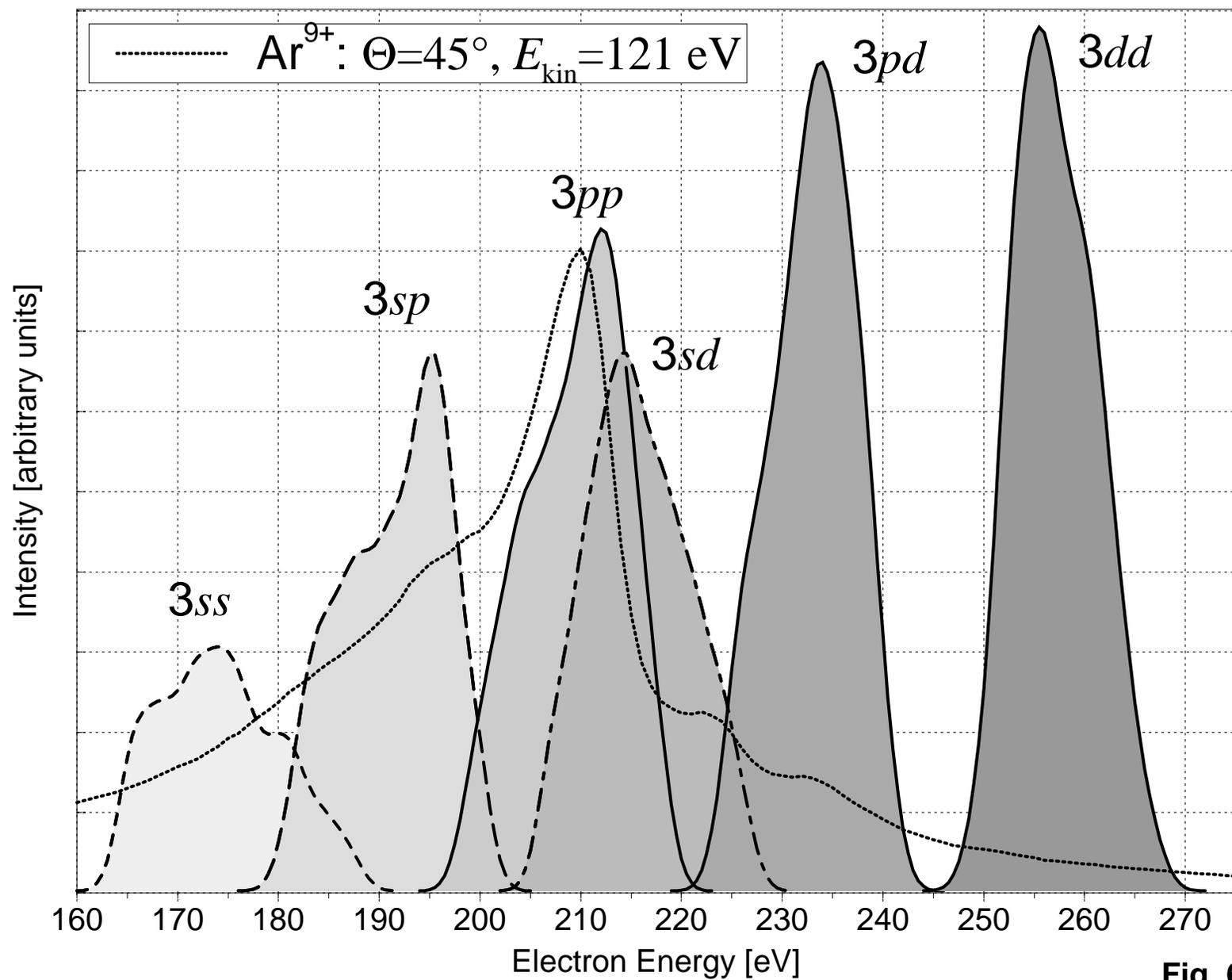
\begin{figure}[p]  
  \caption{This histogram displays the energetic positions
    of $LMM$ Auger transitions originating from neutral ``hollow''
    Ar$^{9+}$ atoms comprising all possibles
    ($2p^53s^xp^yd^z$)-configurations with $n_M=x+y+z \leq 9$.
    Angular coupling shifts have been neglected.  The neutralization
    for configurations possessing less than $q=9$ electrons in the
    $M$-shell has been established via ($q-n_M$) ``spectator''
    electrons in the $N$-shell. Each transition has been weighted by
    unity discarding transition rates and statistical factors due to
    different subshell occupations.}
  \label{fig:Ar9:LMM:histo:spec}
\end{figure}

Let us now look at the grouping of $LMM$ transitions which is plotted in
Fig.~\ref{fig:Ar9:LMM:histo:spec}. The histogram displays the
energetic positions of all $LMM$ transitions originating from initial
$2p^53s^xp^yd^z$ configurations ($n_M=x+y+z \leq 9$) of ``hollow''
Ar$^{9+}$ atoms which are neutralized via $q-n_M$ ``spectator''
electrons in the $N$-shell. Angular momentum coupling as in
\cite{Sch94} is not taken into account. Each transition is weighted by
unity in the plot discarding transition rates and statistical factors
due to different subshell occupations.
For the sake of clarity, the whole spectrum is convoluted by a
Gaussian function of constant width 2eV. This modification evens out
conglomerations of Auger lines at certain energies which are an
artifact of strictly applying the spectator electron approximation.
The width is sufficiently small not to lead to an additional overlap
of $LMM$ subgroup intensities.

Within the same group, Auger transition energies generally tend to
increase steadily with the overall shell population. For comparison,
the dotted line in Fig.~\ref{fig:Ar9:LMM:histo:spec} represents an
autoionization spectrum of Ar$^{9+}$ ions impinging on a Si(100)
surface at $E_{\mbox{kin}}=121$~eV and $\Theta=\Psi=45^\circ$ as reproduced
from the experimental data in
Fig.~\ref{fig:Ar9Si:45deg:energy:normreg}.

Fig.~\ref{fig:Ar9:LMM:histo:spec} reveals that $LMM$ Auger
transitions involving a free and initially neutral Ar atom can cover
the energy interval between 166eV ($2p^53s^24s^2p^5 \mapsto
2p^63s^04s^2p^5$) and 267eV ($2p^53d^9 \mapsto 2p^63d^7$). For
convenience, the groups of $LMM$ transitions displayed in
Fig.~\ref{fig:Ar9:LMM:histo:spec} and the following part of the paper
are classified by the angular $\ell$ quantum numbers of two
participating $M$-shell electrons. In all cases, the final
states are made up of the atomic $2p$ level, the remaining {\it
  M\/}-core states and an appropriate continuum state. For $LMM$
processes, we omit the $2p$ level in our notation.

The low-energy part of the $LMM$ spectrum can be assigned to
$3ss$- and $3sp$ transitions. The higher $3sp$ intensity can be
explained by their statistical weight and their $3p$ contribution
clearly enhancing the transition rates. The fact that the two small
peaks arising in some spectra between 190eV and 200eV fall into the
$3sp$ peak region in Fig.~\ref{fig:Ar9:LMM:histo:spec} might be
fortuitous. Due to our coarse resolution concerning the energetic
grouping, we are not able to ascribe these peaks to particular $3sp$
transitions.

Several things indicate that the dominant peak region around 211eV is
composed of $3pp$ transitions out of a massively occupied {\it
  M\/}-shell instead of $3sd$ transitions the energy range of which
also covers this peak region. At first, it is intuitively plausible,
considering that all three bound state wave functions possess the same
angular momentum, that the by far highest $LMM$ rates are
calculated for the $3pp$ group.  Second, the sharp upper edge of the
211eV-maximum resembles the upper boundary of the $3pp$ curve which is
composed of $3pp$ transitions out of a completed $M$-shell. Due
to level filling statistics, a sharp edge is unlikely to form if its
corresponding transitions take place out of intermediate shell
occupations.  Third, atomic structure calculations yield that $3pp$
energies accumulate around 211eV for all initial $3s^2p^yd^z$
configurations ($y+z \geq 5$), regardless of the particular choice of
$y$ and $z$.  This automatically implies that prior to the majority of
all $3pp$ decays either more than seven electrons have to be captured
into the $M$-shell or the induced charge cloud provides an
equivalent screening effect.

According to the $LMM$ grouping in
Fig.~\ref{fig:Ar9:LMM:histo:spec} we can assign the two subpeaks on
the high-energy shoulder of the $LMM$-maximum to $3sd$- and
$3pd$ transitions, respectively.  $3pp$ processes are unlikely to
contribute to the region above 213eV since they require at least one
$3s$ vacancy along with a ninefold occupied $M$-shell.  These
initial configurations will immediately be converted into $3s^2$
configuration due to the very fast super Coster-Kronig (sCK) decay
channel involving three $M$-shell electron levels.

The spectral range of the $3pd$ peak is cut off at about 235eV and
$3dd$ transitions do obviously not produce enough intensity to appear
with a distinct peak region in the spectra.  These observations
provide experimental evidence that the $3d$ level cannot be
completely populated within the bulk and that quick sCK transitions
tend to carry $3d$ populations into lower lying sublevels before
$LMM$ transitions take place. The missing structures and the spectral
range of the high-energy tail extending above 300eV suggest that it
consists of the large variety of LXY transitions with X,Y$\in$\{N, O\}
rather than $3dd$ transitions.

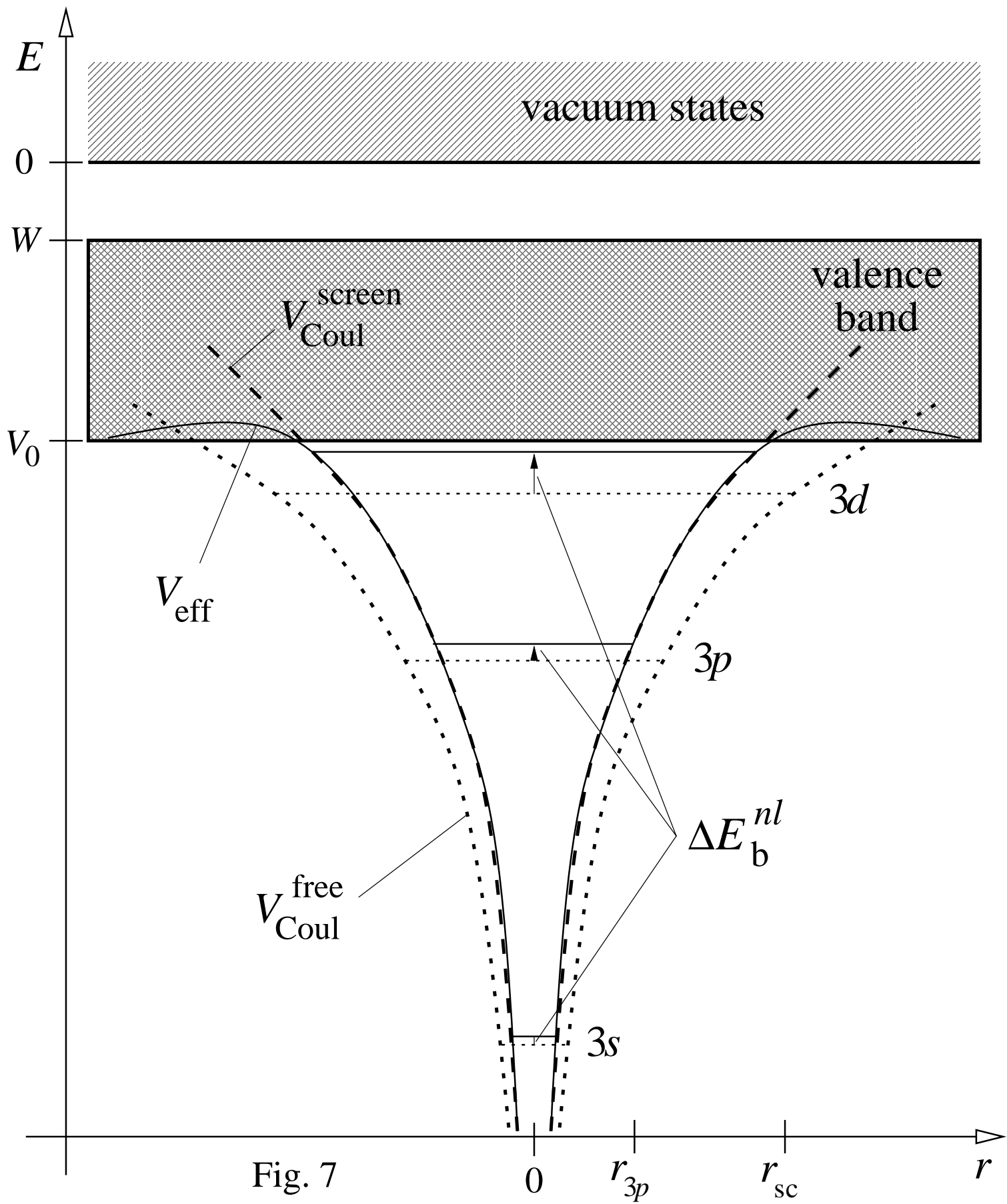
\begin{figure}[p]  
  \caption{Potential $V_{\mbox{eff}}$ (schematic) induced by an HCI core
    which is embedded into an electron gas. $V_{Coul}^{free}$ denotes
    a free ionic Coulomb potential. By adding outer spectator
    electrons to the atom, a screened Coulomb potential
    $V_{Coul}^{screen}$ forms which is implemented in our Cowan code
    calculations (spectator electron model). At large distances from
    the nucleus, the potential $V_{\mbox{eff}}$ seen by an active
    electron merges into the bottom of the valence band $V_0$.
    Electrons occupying states above $V_0$ cannot be bound by the
    ionic core.  They are lost to the valence band continuum.  As
    level binding energies $E_b^{nl}$ approach $V_0$, level shifts
    $\Delta E_b^{nl}$ increasingly deviate from the spectator electron
    model.}
  \label{fig:potential}
\end{figure}

The $LMM$ cut-off at 235eV can be understood by taking a deeper look at
the effective projectile potential $V_{\mbox{eff}}$ within the bulk (see
Fig.~\ref{fig:potential}) which is deformed with respect to the
corresponding free ionic Coulomb potential $V^{free}_{Coul}$.  Close
to the projectile nucleus $r \ll a_0$, the effective potential
$V_{\mbox{eff}}$ converges into $V^{free}_{Coul}$. At intermediate distances
$r \simeq a_0$, the screening of outer levels and the electron gas
starts to act on the projectile levels. In this domain $V_{\mbox{eff}}$ is
well represented by a free atom potential $V^{screen}_{Coul}$ which is
screened by outer shell spectator electrons. All $nl$ subshells with
energies $E^{nl}_b$ are elevated by a subshell dependent amount of
$\Delta E^{nl}_b$ with respect to $V^{free}_{Coul}$. Far away from the
nucleus the effective potential $V_{\mbox{eff}}$ merges into $V_0$ denoting
the bottom of the valence band.

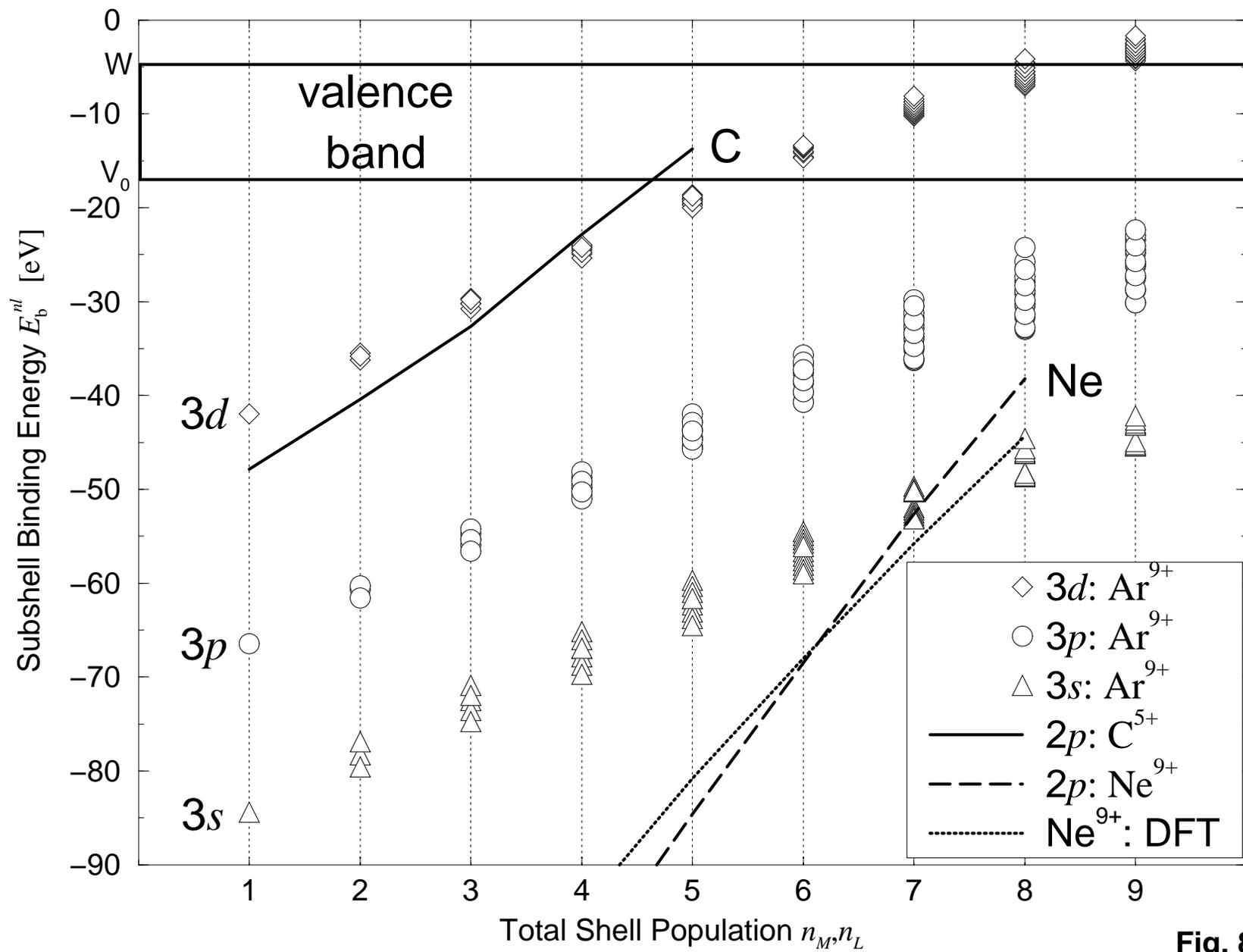
\begin{figure}[p]  
  \caption{Binding energy ranges of hollow atom
    Ar$^{9+}$ $3\ell$ subshells as a function of the total $M$-shell
    population. Within the spectator electron model, the $3d$ shell
    moves above the bottom of the valence band $V_0$ as soon as more
    than five electrons populate this level.  For comparison, we added
    the $2p$ level binding energies of C$^{5+}$ and Ne$^{9+}$ as a
    function of the total $L$-population $n_L$ and a DFT calculation
    \protect\cite{Arn95} for the Ne$^{9+}$--Al system.}
  \label{fig:bind}
\end{figure}

Fig.~\ref{fig:bind} displays the $M$-sublevel binding energies
$E^{nl}_b$ of Ar$^{9+}$ as a function of the total $M$-shell population
$n_M$. The values have been calculated by the Cowan code for spectator
electron configurations, i.e., for the potential $V^{screen}_{Coul}$.
This modeling has proven to yield good agreement with experimental and
more sophisticated theoretical results in the past. In a work by
Schippers \textit{et al.}~\cite{Sch94}, the main $KLL$ peak energies of
the hydrogenlike second row ions C$^{5+}$, N$^{6+}$, O$^{7+}$,
F$^{8+}$ and Ne$^{9+}$ have been reproduced. Arnau \textit{et al.}
\cite{Arn95} have demonstrated that the spectator electron model
complies with DFT calculations including nonlinear screening effects
for hydrogenlike Ne$^{9+}$ ions in an Al target. Detailed calculations
even reveal that the induced charge density tries to mimic the shape
of the wavefunctions of the neighboring unoccupied atomic level.

In Fig.~\ref{fig:bind} we added $2p$ binding energies of hydrogenlike
C$^{5+}$ and Ne$^{9+}$ as obtained from the spectator model and for
comparison the DFT calculation for Ne$^{9+}$ as a function of the
total $L$-shell population $n_L$. Following \cite{Arn95}, the screening
of the atomic spectator electrons resembles the screening by the VB
electron gas because the inner atomic levels are energetically
separated from the VB much like they are separated from next higher
subshell in a free atom. This argument holds for the Ar$^{9+}$ $3s$-
and $3p$ level and also for nearly all $L$-shell levels in
hydrogenlike HCIs which are situated between the C$^{5+}$ and
Ne$^{9+}$ curves.

The evolution of the $3d$ sublevel energies with $n_M$ in
Fig.~\ref{fig:bind} differs from the lower lying subshells, though.
We observe that the $3d$ level binding energies are significantly
closer to the VB and grow above $V_0$ as soon as more than five
electrons populate the $M$-shell. We performed a DFT calculation
showing that $3d$ electrons are already lost to the VB continuum for
$n_M>4$. The spectral cut-off in the $3pd$ transition domain in
Fig.~\ref{fig:Ar9:LMM:histo:spec} can now be explained by omitting all
contributions from $3pd$ transitions with $n_M>4$.  Aiming to correct
for the shape of $V_{\mbox{eff}}$ which largely deviates from
$V_{Coul}^{screen}$ for $E_b^{n\ell} \simeq V_0$ (see
Figs.~\ref{fig:potential} and~\ref{fig:bind}), we shift the atomic
$3d$ level to $V_0$ for $n_M \leq 4$ to attain higher transition
energies compared to the mere spectator electron model. In this manner
we derive the experimental $3sd$- and $3pd$ peak positions on the
high-energy shoulder within an accuracy of 2\% and 1\%, respectively.

\section{Monte Carlo simulation of the subsurface interaction phase}
\label{sec:subsim}

In order to elucidate the interaction mechanism which eventually
generates the measured spectra we worked out a Monte Carlo
simulation~\cite{Kal86}. Our goal was to reproduce the intensity
shifts of the observed spectra for different incident energies in
Fig.~\ref{fig:Ar9Si:45deg:energy:normreg}. On the analogy of previous
simulations by Schippers et al.~\cite{Sch94}, Page et al.~\cite{Pag95}
and Stolterfoht et al.~\cite{Sto95} on the $L$-shell filling of
hydrogenlike highly charged ions at metal surfaces, we only keep track
of the populations of the two innermost projectile shells containing
at least one vacancy and focus on the most dominant transition rates.
The ionic cores are neutralized by $N$-shell spectator electrons. Among
all intra-atomic Auger processes, only those yielding an electron above
the vacuum level are considered.

During the simulation, the three $M$-subshell populations are recorded
continuously. Transition rates, transition energies and sublevel
energies are evaluated dynamically at each iteration step according to
the particular $\{n_{3s}|n_{3p}|n_{3d}\}$ configuration.  From one
step to the next, only the fastest transition which is derived
statistically from its nominal rate takes place. The Monte Carlo
method implies the averaging of the simulation results over a
sufficient amount of projectiles. We find that the simulated spectra
converge after $N \simeq 1 \times 10^5$ particle runs and chose $N=1
\times 10^6$. In our implementation of the subsurface cascade, each
particle is started at the first bulk layer with a fixed angle of
incidence $\Theta=45^\circ$ and energy $E_{\mbox{kin}}$.  For $E_{\mbox{kin}}=121$~eV
and 2~keV we assume an initially empty $M$-shell.

\subsection{Intra-Atomic rates}
\label{sec:intraatomic}

The $LMM$ rates are evaluated by a fit expression proposed by Larkins
\cite{Lar71C} for free multiply ionized atoms possessing no $N$-shell
spectator electrons.  Accordingly, if one or two of the $n$ electrons
of a subshell which could contain $n_0$ electrons are involved in an
Auger process, the Auger rate calculated using the formulae
appropriate for a filled shell $\Gamma^{\mbox{filled}}_{\ell_1 \ell_2}$ is
reduced by $n/n_0$ or $[n(n-1)]/[n_0(n_0-1)]$, respectively.  Values
for $\Gamma^{\mbox{filled}}_{\ell_1 \ell_2}$ are only supplied for $3ss$-,
$3sp$- and $3pp$ transitions in the literature which account for the
greatest part of the overall $LMM$ intensity in the literature.  For
$3sd$-, $3pd$- and $3dd$ transitions, we scale the $LMM$ rates
$\Gamma^{\mbox{filled}}_{3\ell d}$ to reproduce the experimental peak
heights. Table~\ref{tab:intrarates} lists the six
$\Gamma^{\mbox{filled}}_{\ell_1 \ell_2}$ rates which are held constant for
different simulations.

These $LMM$ rates should not be greatly affected by the embedding of the
HCI into the electron gas because they chiefly depend on the radii of
the participating $M$-subshells which remain fairly unchanged. To show
this we recall that the shape of the induced charge cloud is similar
to the $N$-shell. Within the hydrogen atom approximation, the radii of
the screening cloud $r_{sc}$ and the atomic shells (schematically
inserted in Fig.~\ref{fig:potential}) both scale with $(n-1)^2
\{1+\frac{1}{2}[1-\frac{\ell(\ell+1)}{(n-1)^2}]\}$. The ratio
$r_{sc}/r_{3p} = 2.5$ with $sc = 4p$ has to be related to the ratio
$r_{3p}/r_{3s}$ amounting to 0.83. Due to its great extension, the
screening electron cloud should therefore have a minor impact on the
$M$-shell orbitals and hence on the $LMM$ rates given in
Table~\ref{tab:intrarates}.

Since we do not resolve $N$-sublevels, Coster-Kronig $MMN$ transitions
have to be handled by a global base rate for each $M$-level pair.  In a
simple approach, we weight each $MMN$ base rate by the initial
$M$-sublevel occupation and final state vacancies such that the average
rate amounts to $3 \times 10^{14}$s$^{-1}$. For the purposes of this
paper, only the order of magnitude with respect to the other
transition types matters.  We remark that Armen and
Larkins~\cite{Arm91} have calculated transition rates for $MMN$ decay
channels which are of the order of $4 \times 10^{14}$s$^{-1}$,
depending strongly on the angular coupling.  This is in sufficiently
good agreement with our assumption. Only $MMN$ transitions with a final
state above the continuum level are included leaving over solely
$(3s)(3d)$N transitions which are of particular importance for the
initial phase of the interaction.

The sCK $MMM$ rates are known to be 10 to 100 times faster than any
rates for Auger transitions possessing the initial and final holes in
different principal shells. In our simulation, they mainly serve to
regroup any $M$-shell configuration into the appropriate $M$-shell
ground state before $LMM$ transitions take place.  To achieve this, we
utilize a base rate of $1 \times 10^{15}$s$^{-1}$ which is scaled by
the $M$-subshell occupation statistics.  In Table~\ref{tab:sCKrates} we
put together the average number of $MMM$ processes per particle and the
average $M$-sublevel occupation at the time of $MMM$ emission for the
two sCK transitions which are relevant for our simulation. 

\subsection{$MCV$ filling within the bulk}
\label{sec:filling}

Target levels below $V_0$ can be filled by transitions involving
electrons of valence band states (C) which are perturbed by the ionic
core. The energy gain is conveyed either onto another VB electron
which is emitted into the continuum or a collective excitation
(plasmon) is created in the medium.  The theoretical approach
including the charge displacement in the description of the excited
outgoing electrons is much more complicated and, at present, only
unperturbed valence band states (V) are included in the
calculations\cite{Die95,Die96}. The VB electrons take on the role of
outer shells in a free atom.

Using DFT to describe the interaction between the ion and the metal
valence band and following the same scheme as in \cite{Die96}, we have
derived $MCV$ rates for the Ar$^{9+}$--Si system.
Table~\ref{tab:MCVrates} lists the rates per spin state
$\Gamma^{MCV}_{3\ell}$ into the three $M$-sublevels with the number of
initial $M$-shell electrons $n_M$ as parameter.  These $MCV$ rates still
have to be multiplied by the number of unoccupied final states in the
particular $M$-sublevel to attain actual transition rates between two
atomic configurations.  $\Gamma^{MCV}_{tot}$ denotes the overall $MCV$
rate into the $M$-shell after carrying out the appropriate statistics.
Since sCK transitions are much faster than $MCV$s (cf.\ 
Table~\ref{tab:sCKrates}), we only consider ``Coster-Kronig final
states'' as initial configurations in the DFT calculation. The
transition rates are independent of the projectile velocity $v_p$
equaling their static values for all incident energies occurring in
this work.

Table~\ref{tab:MCVrates} reveals that $\Gamma^{MCV}_{3d}$ assumes by
far the highest values. Taking into account the high degeneracy of the
$3d$ level, effective rates $\Gamma^{MCV}_{3d}$ exceed
$\Gamma^{MCV}_{3p}$ and $\Gamma^{MCV}_{3s}$ by more than one and two
orders of magnitude, respectively.  With increasing $n_M$, $MCV$
transfer into the $3p$ state accelerates reaching the
$\Gamma^{MCV}_{3d}$ values at low $n_M$.  This is important
considering that for $n_M>4$ the $3d$ shell vanishes and $MCV$s into
the $3p$ level constitute the most effective $M$-shell filling
mechanism which is eventually responsible for the formation of the
dominant 211eV-peak.

\subsection{Collisional filling}
\label{sec:collisions}

For projectile energies above 1~keV, sidefeeding into the HCI $M$-shell
due to direct electron transfer from target atom core levels supplies
a velocity dependent filling rate.  The transfer crossection increases
with the energetic vicinity of inner projectile and target states
\cite{Gre95} which is maximum for the Ar$^{9+}$ $3s$ level with the
$2p$ bulk level of Si possessing $E_b^{2p}=109$~eV (cf.\ 
Fig~\ref{fig:bind}). Experimentally, a Si target $LMM$ Auger peak for
spectra with $E_{\mbox{kin}} \geq 1$~keV can be observed which is directly
connected to the vacancy transfer. For 2~keV projectiles traveling
through a silicon crystal in (100)-direction, collisional filling
supplies a $3s$ sidefeeding rate of $\Gamma_{3s}^{coll} = v_p/d = 1.8
\times 10^{14}$s$^{-1}$ going on the assumption of one electron
transfer per collision. Within the energy range below 1~keV, collision
frequencies are small and the distance of closest approach is too
large even for head-on collisions to allow a sufficient level
crossing for sidefeeding \cite{Gre95}.

\subsection{Simulation of the 121eV-  and 2~keV-spectra}
\label{sec:121eVspectrum}

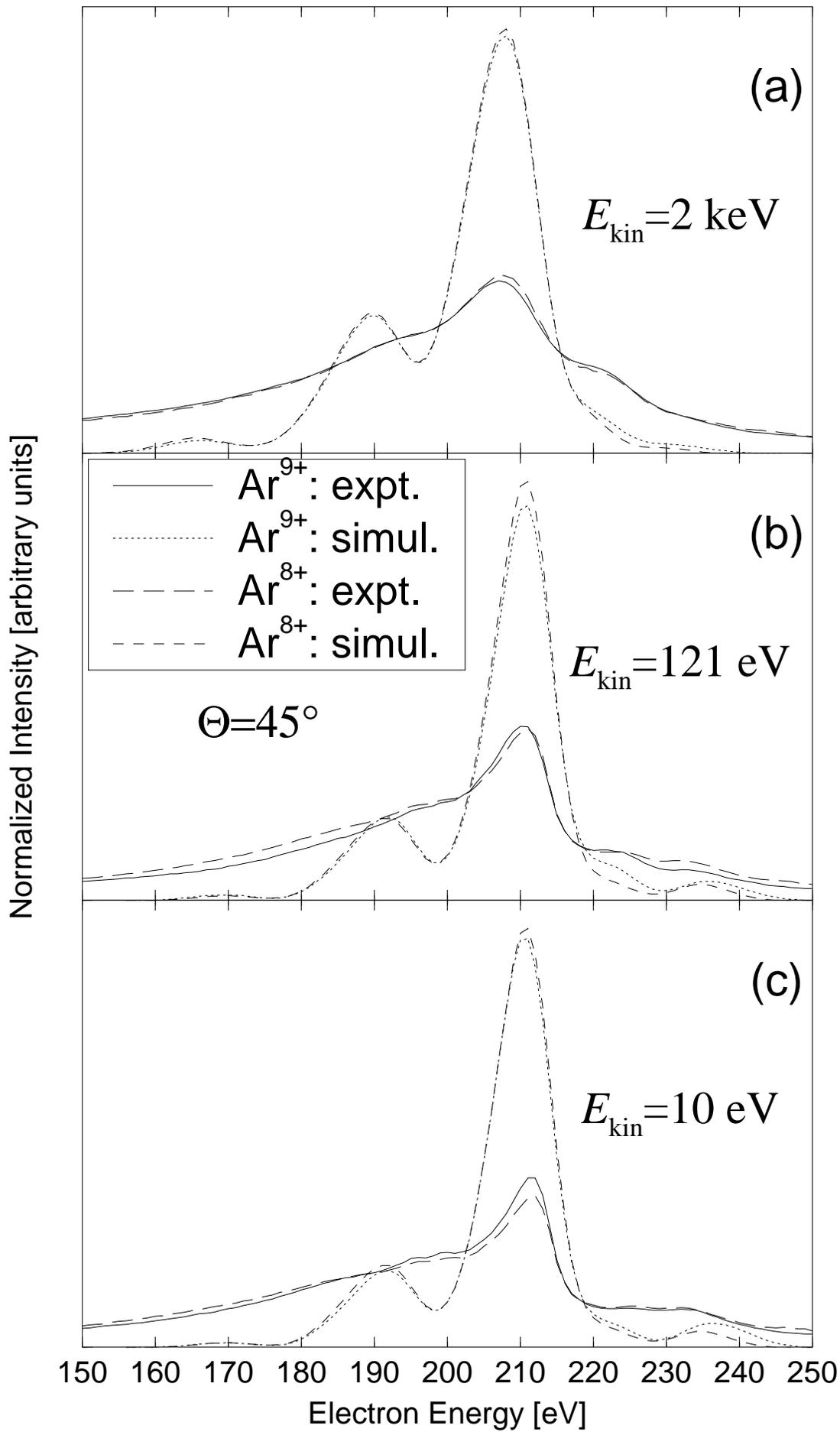
\begin{figure}[p]  
  \caption{Simulation of the experimental   (cf.\
    Fig.~\ref{fig:Ar9Si:45deg:energy:normreg}) Ar$^{9+}$ and Ar$^{9+}$
    spectra at $\Theta=45^\circ$ for $E_{\mbox{kin}}=2$~keV (a) 121eV
    (b) and 10eV (c).  For the ($E_{\mbox{kin}}=10$~eV)-simulation, a
    statistically distributed $M$-shell population with
    $\left<n_M\right>=2$ at the bulk entrance (see text) is
    implemented. The intensity ratios between the $LMM$ transition
    subgroups as a function of $E_{\mbox{kin}}$ display the same
    systematics as the experiment. For the $E_{\mbox{kin}}=2$~keV, we
    enforce a correction for elastic and inelastic bulk damping by
    convolving the original spectrum with an exponential function of
    decay length of 3 a.u..}
  \label{fig:Ar:8:9:Si:sim:exp}
\end{figure}

In Fig.~\ref{fig:Ar:8:9:Si:sim:exp} we plot the experimental spectra
from Fig.~\ref{fig:Ar9Si:45deg:energy:normreg} into three subplots and
compare them with our simulation results which are convoluted by a
Gaussian function of 3eV width. In this section we look at the
Ar$^{9+}$ spectra and postpone the discussion of the Ar$^{8+}$ spectra
which are displayed in the same plot to Section~\ref{sec:Ar8spectra}.
The difference between the simulated spectra in (a) and (b) stems from
collisional filling which is exclusively enabled for $E_{\mbox{kin}}=2$~keV.
In addition, we performed a convolution of the 2~keV-spectrum with an
exponential function with a decay length of 3 a.u.\ to compensate for
elastic and inelastic energy losses of electrons on their way through
the bulk region. For $E_{\mbox{kin}} < 2$~keV, this damping becomes negligible
due to the shallow projectile penetration.

The intensity ratios among the different $LMM$ subgroups and
their peak positions are approximately reproduced. The $3pp$ region
displays too much intensity, though which might be caused by the {\it
  LMM\/} rate fit formula (cf.~Section~\ref{sec:intraatomic})
overestimating the $3pp$ rates for high $M$-populations, see
also \cite{Lar71C}~(Table~VI). The $3ss$ intensity is clearly too low
suggesting that other transitions types not considered in our model
may contribute to this region. The enhancement of the $3sd$ peak
parallel to the disappearance of the $3pd$ peak and intensity gain of
the $3sp$ region towards the $E_{\mbox{kin}}=2$~keV-spectrum as a consequence
of the collisional filling can nicely be observed (cf.\ 
Table~\ref{tab:intrarates}). The average $M$-sublevel
populations at the time of $LMM$ emission (cf.\ 
Table~\ref{tab:intrarates}) indicate that the high-energy shoulder is
generated along the early subsurface interaction phase. On the other
hand, the dominant $3pp$ peak occurs at high $M$-populations
benefitting from the growing $MCV$ rates into the $3p$ level and
the disappearance of the $3d$ level towards high $n_M$. The missing
$3dd$ intensity confirms the presence of the fast $MMN$ and {\it
  MMM\/} decay channels which inhibit the buildup of $3d$ populations
larger than one.

In the experimental spectra, the low-energy tail displays much less
structure than the simulation indicating that the mere spectator
electron model might be incomplete. We carried out other simulations
where 20\% of the $LMM$ transitions start out from singly
ionized initial configurations such that the peak regions loose part
of their intensity to the low-energy side. Doing so the intensity dip
around 200eV gets partially ironed out and the low-energy tail
stretches beyond 160eV. A similar effect could be induced by the
consideration of L$_{2,3}MMM$ double Auger processes
\cite{Abe75} for which Carlson and Krause \cite{Car65} measured a
relative contribution to all radiationless transitions of 10$\pm$2\%
and energy shifts of more than 10eV \cite{Sie69}. For the sake of the
clarity of the displayed simulation results we did not implement this
correction in Fig.~\ref{fig:Ar:8:9:Si:sim:exp}.

\subsection{Simulation for a statistical initial $M$-population}
\label{sec:simulation}

It is very surprising that by reducing the incident energy from about
121eV to 9eV, a significant shift in the relative peak intensities
still takes place. On the one hand velocity dependent below-surface
filling can be ruled out in this energy domain, on the other hand this
effect must originate from different subshell populations at the time
of $LMM$ emission. Let us assume for the moment that individual
$M$-subshells of each particle are filled statistically (by a
Poisson distribution which is cut off at the subshell degeneracy) at
the first bulk layer according to their respective degeneracy, i.e.,
$\left<n_{3\ell}\right>$=2/18, 6/18 and 10/18, multiplied by the mean
total $M$-shell population $\left<n_M\right>$ for $\ell=3s$-,
$3p$- and $3d$ level, respectively. In
Fig.~\ref{fig:Ar:8:9:Si:sim:exp} we present results of a Monte Carlo
simulation with $\left<n_M\right>=2$.

For a greater part of these initial configurations, new {\it
  M\/}-shell redistribution channels open up via $MMN$s and sCKs
which are energetically forbidden for $n_M=0$ and carry part of the
$3d$ population immediately into the $3p$- rather than the $3s$ level.
The simulations in Fig.~\ref{fig:Ar:8:9:Si:sim:exp}(b,c) and
Table~\ref{tab:intrarates} indeed reproduce the intensity shift from
the $3sd$ peak to the $3pd$ peak at 232eV going from
$E_{\mbox{kin}}=$121eV to 9eV. We remark that this simple model of an
initial $M$-shell population before bulk penetration does not
hold exactly for the Ar$^{8+}$ simulation where we set $n_{3s}=1$,
$\left<n_{3p}\right>=1$ and $\left<n_{3d}\right>=1$. We are going to
provide a physical motivation for the model in
Section~\ref{sec:Ar8spectra}.

\section{The evolution of the subsurface cascade}
\label{sec:evolution}

According to the experimental clues and arguments of
Sections~\ref{sec:observations} and~\ref{sec:grouping}, the
overwhelming part of the high-energy branch originates from
below-surface emission. For this phase, we designed the simulation
presented in the previous section. In the following we describe the
evolution of the subsurface cascade on the basis of the simulation
results combined with the experimental data.

As the HCI penetrates into the crystal bulk region all electrons that
have previously been captured into outer Rydberg levels will be lost
and band electrons will neutralize the core charge over a distance of
roughly the Debye screening length of the electron gas. Thus a second
generation of hollow atoms emerges within the bulk.

Prior to any electron capture, the $O$-shell of the Ar$^{9+}$
core is the uppermost ionic shell to still fit below $V_0$. As long as
not more than two electrons populate inner levels, solely XCV
transitions (with X$\in$\{L, M, N, O\}) can proceed.  Since the XCV
transition probability increases with the effective screening and
degeneracy of the final level, XCVs preferably populate the {\it
  O\/}-shell. Before any significant NOO and MNO Auger emission can
take place, the rapid XCV filling successively pushes the $O$-
and $N$-shell above $V_0$.  This period is accompanied by LCV,
LNO, LMN transitions etc.\ creating the smoothly decreasing part of
the spectrum above the $3pd$ edge. We note that this early phase of
the neutralization may already start before complete bulk penetration
when the projectile travels through the vacuum tail of the valence
band.

The loss of whole atomic shells into the valence band stops when the
$M$-shell is reached. At this point of the scenario, a low {\it
  M\/}-shell population with a statistical preference for the
$3d$ level (due to its high degeneracy) is likely to occur. {\it
  MMN\/}-CK processes transfer these $3d$ electrons quickly into the
$3s$ level before a large $3d$ population can accumulate. Other {\it
  MMN\/} transitions $(3p)(3d)$N and later $(3s)(3p)$N are
energetically forbidden. This $M$-shell redistribution is
accelerated by high speed sCK processes with rates of the order of
$10^{15}$s$^{-1}$.  Whereas $3sdd$ transitions are immediately
possible, $MMM$ transitions into the $3p$ level require $n_M >
3$.  Along this early $M$-shell redistribution phase the {\it
  M\/}-population remains fairly constant at $n_M \simeq 2+n_{3s}$,
though because one $M$-electron is lost along each $MMM$
process. Thus $3sd$-$LMM$ processes out of initial
$3s^2d$ constellations are characteristic for this phase causing the
224eV-peak in the experimental spectra. It lasts comparatively long
because the condition $n_M \simeq 2$ keeps the $MCV$ rates
(cf.~Table~\ref{tab:MCVrates}) minimal.

The Ar$^{9+}$ core will always be surrounded by an induced VB charge
cloud (C) because the number of bound states $n_b$ below $V_0$ is
smaller than the projectile core charge $q=9$ (Fig.~\ref{fig:bind}).
Hence $MCV$ processes continue to populate empty $M$-levels faster and
faster with increasing $n_M$.  As soon as $n_M>3$ is satisfied,
$3pdd$ sCKs become energetically possible and a $3p$ population
builds up while the $3d$ population remains approximately at one due
to the presence of the $MMN$  and sCK decay channels.
$3dd$ transitions require the transient formation of very unstable
$M$-shell configuration that are unlikely to occur so they do not
appear in the spectra. At $n_M>4$, the $3d$ level vanishes into the
valence band thus interrupting further $3sd$- and $3pd$ emission.
Since the $3pp$-$LMM$ transitions possess much higher rates than any
other $LMM$ transitions they clearly prevail during this later stage of
the subsurface interaction.

The dominant peak which is centered at 211eV for $E_{\mbox{kin}} \leq 121$~eV
in Fig.~\ref{fig:Ar9Si:45deg:energy:normreg} corresponding to
$3pp$ transitions with $n_M \geq 7$ provides evidence for the
described mechanism, in particular for the high $MCV$ rates into the
$3d$- and later the $3p$ level.  The intensity gain of the
$3sd$ peak with respect to the $3pd$ peak for high $E_{\mbox{kin}}$ is
consistent with the greater time window of the former transition
during the early interaction phase. This effect furthermore verifies
the assumption of collisional sidefeeding into the $3s$ level and
therefore the 224eV-peak assignment by itself. All phases are
accompanied by $3ss$- and $3sp$-$LMM$ transitions which constitute the
low-energy tail and the region around the two faint subpeaks between
about 180eV and 200eV, respectively.

\section{Spectra of metastable A\lowercase{r}$^{8+}$ projectiles}
\label{sec:Ar8spectra}

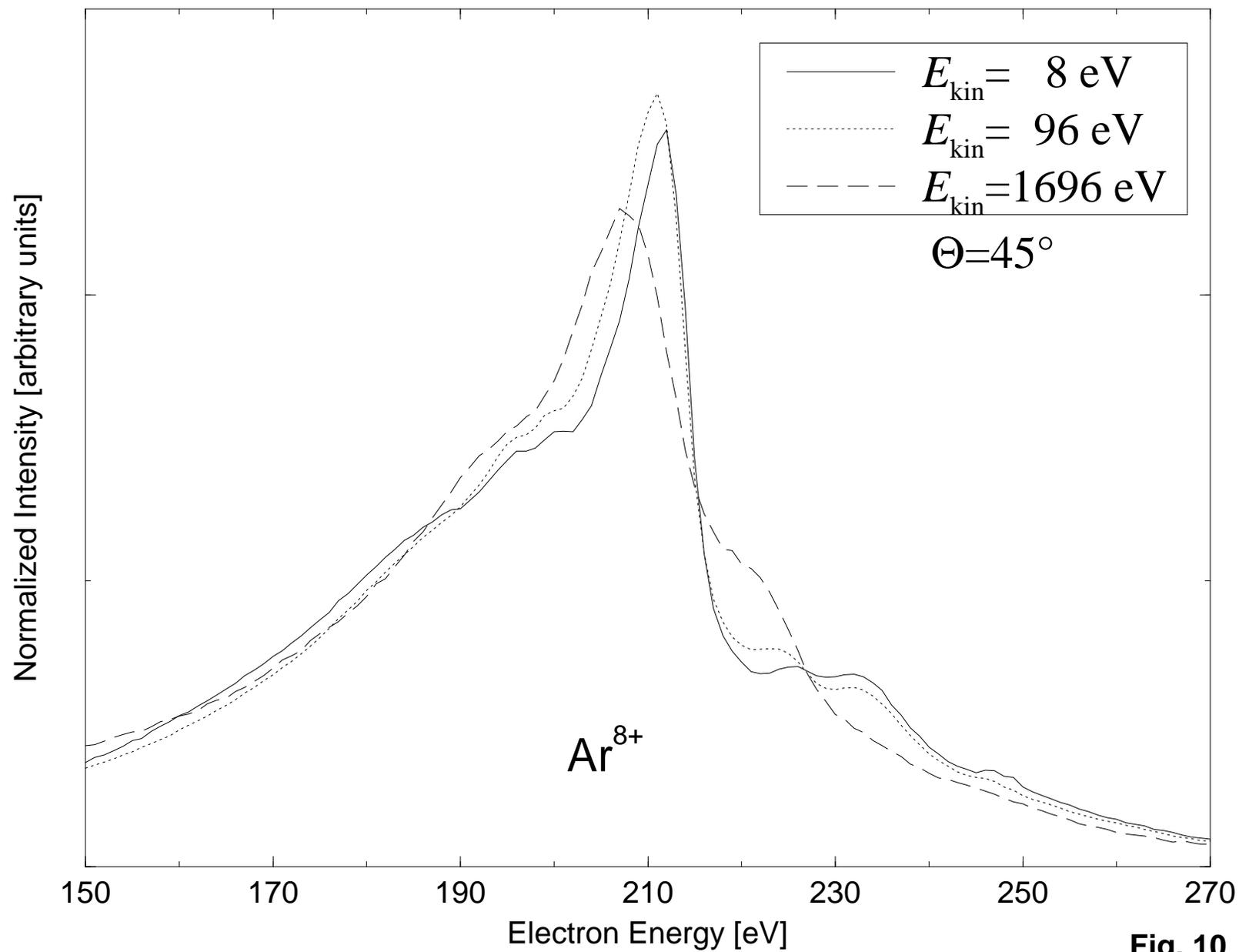
\begin{figure}[p]  
  \caption{Experimental spectra of metastable Ar$^{8+}$ projectiles
    impinging at a fixed incident angle of $\Theta=45^\circ$ and
    varying $E_{\mbox{kin}}$ on a $n$-Si(100) surface. Despite the
    initial $3s$ electron, the general shape of the spectra is very
    similar to the corresponding Ar$^{9+}$ spectra in
    Fig.~\ref{fig:Ar9Si:45deg:energy:normreg}. Only the $3ss$- and
    $3sp$ region intensities are slightly enhanced. The small peak at
    247eV occurs at minimal $v_\perp$ and can be assign to
    above-surface {\it LMV}$_W$ emission \protect\cite{Duc97L}.}
  \label{fig:Ar8Si:45deg:energy:normreg}
\end{figure}

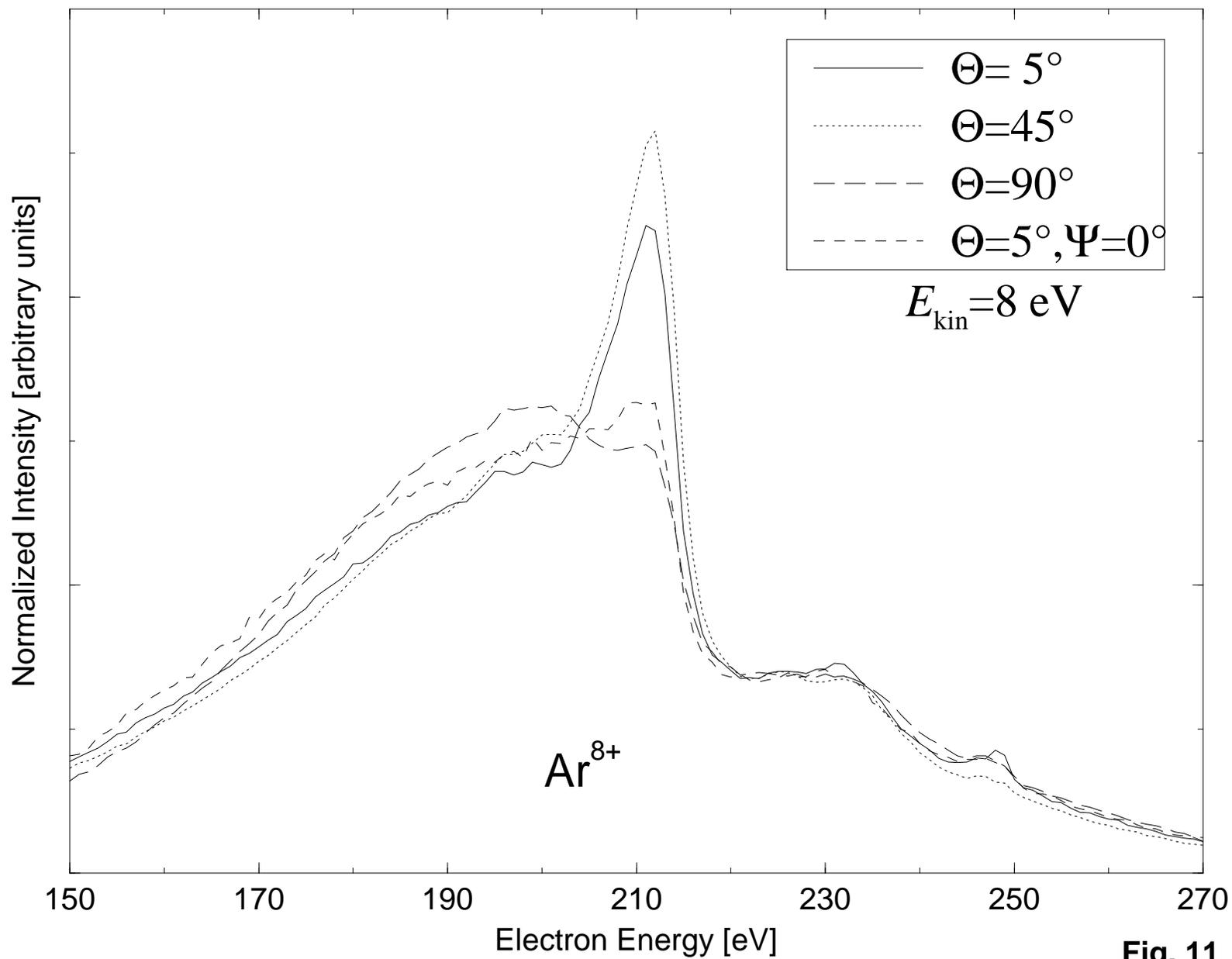
\begin{figure}[p]  
  \caption{Experimental spectra of metastable Ar$^{8+}$ projectiles
    impinging at $E_{\mbox{kin}}=8$~eV and different incident angles
    $\Theta$ on a $n$-Si(100) surface. All spectra exhibit the
    small peak at 247eV as expected for the minimum perpendicular
    projectile velocity components $v_\perp$. At $\Psi = 0^\circ$, the
    broad peak around 198eV pops up again, cf.\ 
    Figs.~\ref{fig:Ar9Si:121eV:angle:normreg}--\ref{fig:Ar9Si:92deg:energy:normreg}.
    In the ($\Theta=5^\circ,\Psi=0^\circ$)-geometry, its intensity is
    reduced with respect to the 211eV-peak as compared to the
    corresponding Ar$^{9+}$ spectrum in
    Fig.~\ref{fig:Ar9Si:121eV:angle:normreg}.}
  \label{fig:Ar8Si:8eV:angle:normreg}
\end{figure}

Seeking to extract additional experimental evidence for the described
Ar$^{9+}$ interaction mechanism, we performed a series of
measurements involving metastable ($2p^53s$) Ar$^{8+}$ ions colliding
at $\Theta=45^\circ$ and various kinetic energies with a Si crystal
(Fig.~\ref{fig:Ar8Si:45deg:energy:normreg}). A straight comparison
with the corresponding Ar$^{9+}$ series in
Fig.~\ref{fig:Ar9Si:45deg:energy:normreg} shows that the general shape
of the spectra is unaffected by the additional $3s$ electron except
for a slight enhancement of the $3ss$- and $3sp$ intensities.  In
fact, the only new structure observed is a small peak arising at 247eV
for $E_{\mbox{kin}}=8$~eV and generally for lowest perpendicular projectile
velocities $v_\perp$ which can also be deduced from
Fig.~\ref{fig:Ar8Si:8eV:angle:normreg}.

The 247eV-peak has been discussed in detail in \cite{Duc97L} along
with corresponding peaks which occur under similar conditions in the
spectra of second row ions in $1s2s$ configurations. It can be
assigned to so called {\it LMV}$_W$ transitions in the course of which the
$3s$ electron jumps into the $2p$ vacancy. The emitted electron
comes from a level possessing a binding energy which equals the target
work function $W=4.6$~eV for silicon.  Due to the shape of the
subsurface potential $V_{\mbox{eff}}$ (Fig.~\ref{fig:potential}), these
levels cannot exist after projectile penetration into the bulk
occurred. As mentioned earlier in Section~\ref{sec:grouping}, the
strong decrease in spectral intensity above 235eV gives evidence for
this assertion.

The identification of an above-surface {\it LMV}$_W$ peak suggests that
inner atomic shells X$\in$\{M, N, O, $\ldots$\} could be partially
filled before bulk penetration by an autoionization process
XV$_W$V$_W$. We mentioned earlier that also $MCV$ set in with
continuously increasing rates as the HCI travels through the vacuum
tail of the valence band.  Compared to the $MCV$ filling within the
bulk, these near-surface $M$-shell filling channels are likely to
proceed significantly slower, though.  Since sCK processes require
certain minimum $M$-shell populations they are widely inhibited for
these constellations. One can thus expect that a very slow projectile
might enter the bulk region with a low $M$-shell population $\left< n_M
\right> \simeq 2$ favoring the $3d$ level due to its degeneracy.
This way we can motivate the ansatz for the simulation of the spectra
at minimum $E_{\mbox{kin}}$ in Section~\ref{sec:simulation}, even though an
explicit experimental evidence is still missing.

The astonishing similarity of the rest of the Ar$^{8+}$  and the
Ar$^{9+}$ data bears out our previous assumption of fast $MCV$, $MMN$
and sCK processes within the bulk redistributing any $M$-shell
population swiftly into the $3s$ level. In order to compensate for
the additional $3s$ electron in Ar$^{8+}$ $M$-shell, sCKs have to
proceed before an $LMM$ transition takes place. This automatically
implies that the $M$-shell must be sufficiently populated and quickly
replenished at this point. Because a large $M$-shell population far in
front of the surface would be in contradiction to all previous
experiments we can exempt the above-surface zone as the origin of the
emitted electrons. This obviously also holds for the $3pd$ peak at
232eV.

We made use of the great correspondence of the Ar$^{8+}$  and the
Ar$^{9+}$ spectra to check the mechanisms and rates entering our
interaction model. For the simulations on Ar$^{8+}$ projectiles which
are also shown in Fig.~\ref{fig:Ar:8:9:Si:sim:exp} we kept the same
transitions types and rates but added an $3s$ electron to the initial
$M$-shell population. Within the accuracy of our interaction model, the
similarity of the two series is well reproduced.

\section{Summary and discussion}
\label{sec:discussion}

In this work we have presented detailed experimental results on the
interaction of Ar$^{9+}$  and metastable Ar$^{8+}$ ions impinging on
a Si(100) crystal. Doing so we focused on autoionization spectra
measured at low impact energies. In this energy domain, we identified
several new spectral features which alter with the perpendicular
projectile velocity component and with the angle of incidence and
observation. A consistent interaction model has been suggested for
which $MCV$ processes and the energetic vicinity of the
Ar$^{9+}$ $3d$ subshell to the bottom of the silicon valence band
play a decisive role.

The subsurface interaction phase has been simulated using a
Monte Carlo code. Feeding the code with realistic transition rates,
we have been able to reconstruct the experimental peak positions and
intensity shifts for different projectile energies. Our results give
indirect evidence for a very effective below-surface $MCV$ filling as
postulated by theory. In contrast to $KLL$ spectra of hydrogenlike
second row ions impinging on metal surfaces, the main intensity of the
Ar$^{9+}$ $LMM$ spectra is located on the high-energy side of the peak
region corresponding to a massively occupied $3p$ subshell. We
demonstrated that this peculiar shape of the high-energy region is
linked to the special role of the $3d$ subshell which mediates a fast
$M$-shell filling in the beginning and later disappears due to the
screening of the valence band electron gas.

We presented spectra measured at small observation angles with respect
to the surface parallel. They contain a high intensity peak region
which most likely originates from Auger emission of incoming or
reflected projectiles which do not experience the full bulk screening,
yet. In addition, we spotted a distinct peak in the Ar$^{8+}$ spectra
for the lowest perpendicular incident velocities which can be
explained by a unique above-surface process involving the $L$-vacancy
and two electrons from the resonantly populated shells. 

HCI beams have been deemed a candidate for future surface modification
techniques for some time. It has been demonstrated that single ions
can give rise to nanoscale size features on certain surfaces
\cite{Par95}. Also sputter yields on insulators could be significantly
enhanced by using slow HCIs instead of fast singly charged
projectiles.  At very low kinetic energies, the energy deposition
concentrates on a very small area which extends approximately one
lattice constant in the vicinity of the first bulk layer. In this
manner, an energy of several keV can be carried into this zone where
it might be converted into activation energy for processes like
sputtering, crystal growth and surface catalysis. Research in this
field is under way and first results have been presented already.

\section*{Acknowledgments}

This work was sponsored by the German Bundesministerium f\"ur Bildung,
Wissenschaft, Forschung und Technologie under Contract No. 13N6776/4.
We are also grateful for support from the Ministerium f\"ur
Wissenschaft und Forschung des Landes Nordrhein-Westfalen.


\newpage


\pagebreak


\begin{figure}[p]  
  \caption{Target chamber geometry. (a) The HCI beam collides with the
    target surface at an incident angle $\Theta$. Electrons are
    emitted at an observation angle $\Psi$ with respect to the target
    surface towards the spectrometer entrance lens. In most
    experiments, we chose $\Theta+\Psi=90^\circ$. For $\Psi \gg
    0^\circ$, electron emission originating from all interaction
    phases can be detected. (b) At grazing observation $\Psi =
    0^\circ$, the detection of electrons emitted from below the
    surface is suppressed due to their damping on the long path
    through the bulk towards the spectrometer lens.  Only electrons
    emitted from the above-surface zone are not affected such that
    spectral regions stemming from transitions of incoming or
    reflected projectiles gain intensity with respect to regions
    ascribed to below-surface emission.}
  \label{fig:geometry}
\end{figure}

\begin{table}[p]
  \begin{center}
    \leavevmode
    \begin{tabular}{|c|c|c|c|c|} \hline
      process & $\Gamma^{\mbox{filled}}_{3\ell_1\ell_2}$ & $E_{\mbox{kin}}=9$~eV ($\left<n_M \right>=2$) & $E_{\mbox{kin}}=121$~eV & $E_{\mbox{kin}}=2$~keV \\ \hline
      $3ss$ & $3.31 \times 10^{12}$ &  0.8\% (2.0$|$4.5$|$0.1) &  1.0\% (2.0$|$4.3$|$0.1) &  2.0\% (2.0$|$4.1$|$0.1) \\ \hline
      $3sp$ & $5.29 \times 10^{13}$ & 15.9\% (1.6$|$5.1$|$0.0) & 17.1\% (1.7$|$5.1$|$0.0) & 22.1\% (2.0$|$5.0$|$0.0) \\ \hline
      $3pp$ & $1.98 \times 10^{14}$ & 72.5\% (1.4$|$5.5$|$0.0) & 70.0\% (1.5$|$5.5$|$0.0) & 66.8\% (2.0$|$5.4$|$0.0) \\ \hline
      $3sd$ & $6.20 \times 10^{14}$ &  5.3\% (1.2$|$0.5$|$1.4) &  7.2\% (1.2$|$0.4$|$1.5) &  7.4\% (2.0$|$0.3$|$1.4) \\ \hline
      $3pd$ & $1.65 \times 10^{15}$ &  4.6\% (0.6$|$1.5$|$1.4) &  3.7\% (0.8$|$1.4$|$1.4) &  1.4\% (1.5$|$1.2$|$1.2) \\ \hline
      $3dd$ & $4.13 \times 10^{14}$ &  0.8\% (0.4$|$0.4$|$2.4) &  1.0\% (0.5$|$0.2$|$2.3) &  0.4\% (1.2$|$0.1$|$2.2) \\ \hline
    \end{tabular}
    \caption{Monte Carlo simulation results on $LMM$ processes
      for Ar$^{9+}$ impinging on Si(100) with $E_{\mbox{kin}}=9$~eV,
      121eV and 2~keV.  $\Gamma^{\mbox{filled}}_{3\ell_1\ell_2}$ gives
      the $LMM$ rate for a filled $M$-shell as required
      for the implemented fit formula \protect\cite{Lar71C}. For each
      simulation, we list the relative intensity and, in brackets, the
      average ($n_{3s}|n_{3p}|n_{3d}$)-configuration at the time of
      $LMM$ decay which provides information about the evolution
      of the subsurface cascade.}
    \label{tab:intrarates}
  \end{center}
\end{table}

\begin{table}[p]
  \begin{center}
    \leavevmode
    \begin{tabular}{|c|c|c|c|} \hline
      process & $E_{\mbox{kin}}=9$~eV ($\left<n_M \right>=2$) & $E_{\mbox{kin}}=121$~eV & $E_{\mbox{kin}}=2$~keV \\ \hline
      $3sdd$ & 66.2\% (0.2$|$0.4$|$2.4) & 81.8\% (0.2$|$0.2$|$2.4) & 11.7\% (1.0$|$0.1$|$2.2) \\ \hline
      $3pdd$ & 16.3\% (1.0$|$0.2$|$2.8) & 21.0\% (1.0$|$0.2$|$1.5) & 17.9\% (1.5$|$0.1$|$2.4) \\ \hline   
    \end{tabular}
    \caption{Monte Carlo simulation results on $MMM$ processes for
      Ar$^{9+}$ impinging on Si(100) with $E_{\mbox{kin}}=9$~eV, 121eV and
      2~keV. The table lists the average occurrence of each transition
      type and, in brackets, the average $M$-sublevel population at
      the time of $MMM$ emission. Other sCK transitions are
      energetically forbidden.}
    \label{tab:sCKrates}
  \end{center}
\end{table}

\begin{table}[p]
  \begin{center}
    \leavevmode
    \begin{tabular}{|c|c|c|c|c|} \hline
      $n_M$ & $\Gamma^{MCV}_{3s}$ [s$^{-1}$]  & $\Gamma^{MCV}_{3p}$ [s$^{-1}$]   & $\Gamma^{MCV}_{3d}$ [s$^{-1}$]   & $\Gamma^{MCV}_{tot}$ [s$^{-1}$]  \\ \hline
      0 & $9.92 \times 10^{11}$ & $2.07 \times 10^{12}$ & $6.61 \times 10^{13}$ & $8.10 \times 10^{14}$ \\ \hline
      1 & $1.21 \times 10^{13}$ & $2.54 \times 10^{13}$ & $9.18 \times 10^{13}$ & $1.08 \times 10^{15}$ \\ \hline
      2 & -                   & $3.26 \times 10^{13}$ & $1.22 \times 10^{14}$ & $1.41 \times 10^{15}$ \\ \hline
      3 & -                   & $4.46 \times 10^{13}$ & $1.44 \times 10^{14}$ & $1.70 \times 10^{15}$ \\ \hline
      4 & -                   & $6.53 \times 10^{14}$ & -                   & $2.61 \times 10^{15}$ \\ \hline
      5 & -                   & $5.78 \times 10^{14}$ & -                   & $1.74 \times 10^{15}$ \\ \hline
      6 & -                   & $4.48 \times 10^{14}$ & -                   & $9.17 \times 10^{14}$ \\ \hline
      7 & -                   & $3.27 \times 10^{14}$ & -                   & $3.27 \times 10^{14}$ \\ \hline
    \end{tabular}
    \caption{$MCV$ rates for the Ar$^{9+}$/Si system. The table
      lists $MCV$ transition rates per spin state
      $\Gamma^{MCV}_{3\ell}$ for each $M$-sublevel and the overall
      $MCV$ rate $\Gamma^{MCV}_{tot}$ taking into account occupation
      statistics as evaluated by DFT calculations.  $n_M$ gives the
      initial number of $M$-electrons.  The rates refer to initial
      $M$-shell ground state configurations.  For $n_M=0$,
      $MCV$ processes filling the $3d$ level possess by far the
      highest rates. As the subsurface cascade proceeds and $n_M>4$,
      the $3d$ level vanishes and the $MCV$s into the $3p$ level
      rapidly populate the $M$-shell.}
    \label{tab:MCVrates}
  \end{center}
\end{table}

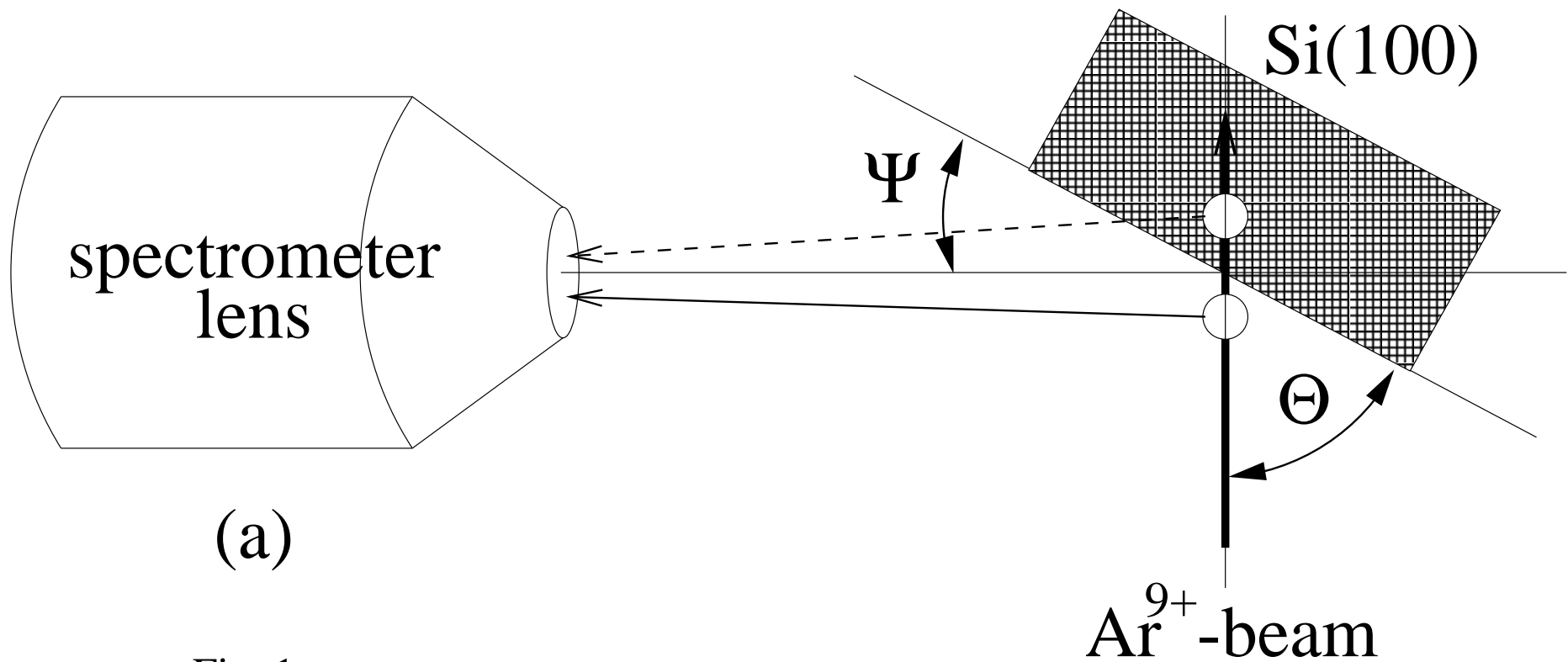
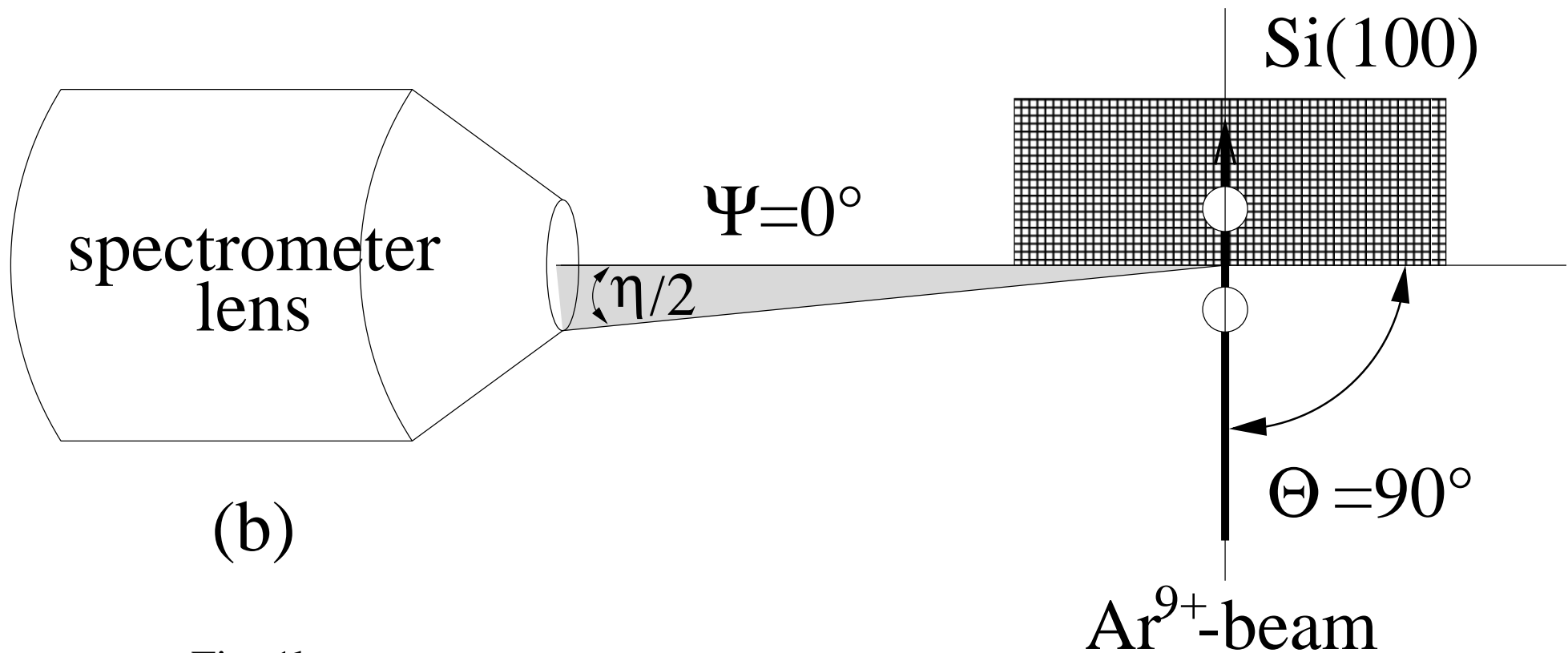
\end{document}